\documentclass{aa}  

\usepackage{graphicx}
\usepackage{txfonts}
\usepackage[amssymb,mediumspace,thickqspace]{SIunits}
\usepackage{subcaption}
\usepackage{amssymb}
\usepackage{float}
\usepackage[labelfont=bf]{caption}

\begin{document}
 
  \title{Characterization of mid-infrared polarization due to scattering in protoplanetary disks}


   \author{S. Heese\inst{1}
          \and
         S. Wolf\inst{1}
         \and 
         R. Brauer\inst{2}
          }

   \institute{Institute of Theoretical Physics and Astrophysics, University of Kiel, Leibnizstraße 15, 24118
Kiel, Germany
\and 
CEA Saclay – Service d’Astrophysique, Orme des Merisiers, Bât 709, 91191 Gif-sur-Yvette, France}

 
  \abstract
  {It is generally assumed that magnetic fields play an important role in the formation and evolution of protoplanetary disks.
  One way of observationally constraining magnetic fields is to measure polarized emission and absorption produced by magnetically aligned
  elongated dust grains.
  The fact that radiation also becomes linearly polarized by light scattering
  at optical to millimeter wavelengths complicates magnetic field studies.}
   {We characterize the linear polarization of mid-infrared radiation due to scattering of the stellar radiation and dust thermal re-emission
   radiation (self-scattering).}
   {We computed the radial polarization profiles
   at wavelengths across the \textit{N} and \textit{Q} bands
   for a broad range of circumstellar disk configurations. These simulations served as a basis to analyze the correlations 
   between selected disk parameters and the resulting linear polarization.}
   {We find that the thermal re-emission radiation is stronger
   than the scattered stellar radiation for disks with inner holes smaller than $\sim10\,\mathrm{au}$ 
   within the considered parameter range.
   The mid-infrared polarization due to scattering shows several clear trends: 
   For scattered stellar radiation only, the linear polarization degree decreases slightly with increasing radial distance, while it
   increases with radial distance for thermal re-emission radiation only and for a 
   combination of scattered stellar radiation and thermal re-emission radiation.
   
   The linear polarization degree decreases with increasing disk flaring and luminosity
   of the central star. An increasing inner
   radius shifts the increase of the linear polarization degree further outside,
   while a larger scale height increases the linear polarization degree for small radial distances and decreases this degree further outside.
   For longer wavelengths, i.e., toward the \textit{Q} band in our study,
   the linear polarization degree converges more slowly.}
   {We found several clear trends for polarization due to scattering. These trends are the basis to distinguish polarization due to
   scattering from polarization due to dichroic emission and absorption.}

   \titlerunning{Characterization of mid-infrared polarization due to scattering}
   \authorrunning{}
   \maketitle
%

\section{Introduction}


It is generally assumed that magnetic fields play an important role in the formation and evolution of protoplanetary disks. 
For example, magnetic fields can cause strong magnetic braking, hindering the formation of a centrifugally supported disk (\citealt{0004-637X-647-1-374}).
A magnetic field oriented vertically with respect to the disk midplane increases angular momentum transport in accretion disks by means of magnetorotational instability
even in ideal magnetohydrodynamics (MHD). In nonideal MHD simulations, the presence of a vertical magnetic field
is important to provide accretion rates that are consistent with observations (\citealt{0004-637X-775-1-73}).
Despite this importance,  magnetic fields in protoplanetary disks are still poorly constrained observationally
(\citealt{2017ApJ...851...55S}).

One way of observationally constraining magnetic fields is to measure polarized emission produced by dichroic emission and absorption
by magnetically aligned elongated dust grains (\citealt{2007ApJ...669.1085C}).
The polarization state of the radiation is described by its linear and circular component, 
where the latter is usually neglected because it is
significantly weaker than the former.
Another established method relies on measuring the Zeeman splitting of spectral lines (\citealt{brauer1}).
By comparing the derivative of the flux profile with the circularly polarized emission, it provides the line-of-sight magnetic field strength (\citealt{1993ApJ...407..175C}).
The total magnetic field can be derived by combining Zeeman observations with complimentary observations of linear polarization
(\citealt{Heiles2012}).
However, the requirements for spatially resolved Zeeman observations cannot be met with current instruments (\citealt{brauer1}).
Only spatially unresolved observations are predicted to be feasible under certain circumstances.

Most polarization observations were done in the submillimeter wavelength regime.
\cite{Rao1} obtained spatially resolved polarization observations of IRAS 16293 - 2422 using the Submillimeter Array.
The magnetic field derived from this observation suggests a complex field structure.
For HL Tau, analysis of spatially resolved polarization observations at $1.25\,\mathrm{\milli\meter}$ (\citealt{2014Natur.514..597S}) shows that
the orientation of the magnetic field is coincident with the major axis of the disk.


Polarization in the mid-infrared (mid-IR) provides an alternative approach to study these fields (\citealt{doi:10.1046/j.1365-8711.2000.03158.x}).
Protoplanetary disks are optically thick in the mid-IR (\citealt{1997ApJ...490..368C}), allowing for the probing of upper disk layers.
This is useful to study the alignment of (sub-) micron sized grains with superparamagnetic inclusions
because their alignment with the magnetic field is predicted to be possible there (\citealt{0004-637X-839-1-56}).
Up to now, there are only a few spatially resolved polarization observations in the mid-IR (e.g., \citealt{2016ApJ...832...18L}). However,
spatially unresolved observations in this wavelength range can be used to constrain the magnetic field as well (\citealt{2018MNRAS.473.1427L}).

However, polarized emission is not produced by dichroic emission and absorption alone, but also by scattering of stellar radiation
and the thermal re-emission radiation of the dust, known as "self-scattering" (\citealt{kataoka1}). 
Spectropolarimetry can help to distinguish dichroic emission from
 absorption (\citealt{2004MNRAS.348..279A, doi:10.1093/mnras/stw2761}), while distinguishing scattering from dichroic
emission and absorption is less straightforward (\citealt{2016MNRAS.460.4109Y}).
Scattering is often neglected in the mid-IR. Yet, mid-IR polarization observations of AB Aur (\citealt{2016ApJ...832...18L})
indicate that both polarization mechanisms are at work: 
dichroic emission and absorption within the inner $\sim 80\,\mathrm{au}$ around the central star 
and scattering at larger radial distances.

In addition, recent follow-up studies of HL Tau have shown that the $870\,\mathrm{\mu\meter}$ emission could arise 
solely from dust scattering (\citealt{2017ApJ...851...55S}), while the emission at $1.3\,\mathrm{\milli\meter}$ shows
signs of both scattering and dichroic emission and absorption. 
Therefore,
a reliable analysis of polarization observations must take scattering into account.


To allow for the distinction of polarization due to scattering from dichroic emission and absorption,
the characterization of the polarization caused by scattering alone 
is necessary.
In the current study we investigate the dependence of the linear, wavelength-dependent polarization degree
on selected parameters describing the central star and disk structure.

This paper is organized as follows: In Sect. 2, the applied Monte Carlo radiative transfer code
as well as the disk and dust setup are described. In Sect. 3, we analyze the results from our simulations.
Afterward, in Sect. 4, we discuss how our results can help to distinguish scattering from dichroic emission and absorption.
Finally, in Sect. 5, we summarize our results.
 
 \section{Setup}
 
 
 In Sect.\,\ref{sec:mcrt}, we describe the applied Monte Carlo radiative transfer code.
 Afterward, in Sect.\,\ref{sec:disk}, we describe the disk model used, before we describe the dust model in Sect.\,\ref{sec:dust}.
 \subsection{Monte Carlo radiative transfer code}
 \label{sec:mcrt}
 
 The simulations presented in this article are produced with the 3D dust radiative transfer code POLARIS 
 (\citealt{2016A&A...593A..87R}\footnote{available online on \url{http://www1.astrophysik.uni-kiel.de/~polaris/}}).
 This code calculates the dust temperature based on the assumption of local thermodynamic equilibrium,
  combining the continuous absorption method
 proposed by \cite{1999A&A...344..282L} and the immediate temperature correction by \cite{2001ApJ...554..615B}.
The code allows the use of multiple polarization mechanisms: first, dichroic emission and absorption by aligned elongated dust grains;
second, scattering of stellar radiation; and third, self-scattering of thermal re-emission radiation.
However, we solely consider scattering by spherical grains as polarization mechanism in this study.

We constructed the polarization maps as follows: In a first step, we computed the temperature structure using a
Monte Carlo approach. In the second step, we computed the scattered stellar radiation and the radiative transfer of the thermal re-emission
radiation based on the temperature profile computed before, taking the effect of self-scattering into account.
 
 \subsection{Disk setup}
 \label{sec:disk}
 
 We applied a parameterized density distribution according to \cite{1973AA....24..337S},
 successfully used in earlier polarization studies (e.g., \citealt{cb26, 2013A&A...553A..69G, doi:10.1093/mnras/stw1692, doi:10.1093/mnras/stx1066}), as follows:
 \begin{align}
  &\rho(r, z)\sim \left(\frac{r}{100\,\mathrm{au}}\right)^{-\alpha}\cdot \exp\left(-\frac{1}{2}\frac{z^2}{h(r)^2}\right), \label{equ:rho} \\
\text{with } &h(r)=h_{100}\cdot \left(\frac{r}{100\,\mathrm{au}}\right)^{\beta}. \label{equ:h}
 \end{align}
In this equation, $r$ is the radial distance from the central star in the disk midplane while $z$ is the height above the midplane.

We used inner disk radii between the dust sublimation radius and those observed in
transition disks (\citealt{doi:10.1146/annurev-astro-081710-102548}).
We considered dust masses ranging from
$10^{-6}\,M_{\odot}$ to $10^{-3}\,M_{\odot}$, to cover the range of masses observed in the
Taurus-Auriga star-forming region (\citealt{0004-637X-631-2-1134}), which spans the widest range of masses of the clusters 
presented in \cite{doi:10.1146/annurev-astro-081710-102548}. 

The outer radii considered cover the range of radii observed in Taurus-Auriga (\citealt{2009ApJ...701..698S}).
The values for the flaring exponent $\beta$, the density distribution exponent $\alpha$, and the 
scale height $h_\text{100}$ are chosen to include the range of values typically found in the literature
(\citealt{2008A&A...489..633P, cb26, 2010A&A...523A..42H, 2012A&A...543A..81M, 2012A&A...546A...7L, 2013A&A...553A..69G, doi:10.1093/mnras/stw1692}).

We considered circumstellar disks around two different central stars, a T Tauri and a Herbig Ae star.
In these systems, light scattering simulations are performed at nine logarithmically distributed wavelengths across the 
\textit{N} and \textit{Q} bands: five wavelengths from $8\,\mathrm{\mu\meter}$
to $13\,\mathrm{\mu\meter}$ and four from $16\,\mathrm{\mu\meter}$ to $22\,\mathrm{\mu\meter}$.
\begin{table}
 \centering
  \begin{tabular}{c c c}
   \hline
    \multicolumn{2}{|c|}{\bf{General}} \\
   \hline
   Distance & $140\,\mathrm{pc}$ \\
   Inclination & $0^\circ - 90^\circ$\\
   Wavelengths & $8\,\mathrm{\mu\meter}-13\,\mathrm{\mu\meter}$ and $16\,\mathrm{\mu\meter}-22\,\mathrm{\mu\meter}$\\
   \hline 
    \multicolumn{2}{|c|}{\bf{Disk}} \\
   \hline
   $\alpha$ & $1.0 - 2.5$ \\
   $\beta$ & $1.05 - 1.4$ \\
   $h_\text{100}$ & $5\,\mathrm{au} - 15\,\mathrm{au}$ \\
   $r_\text{in}$ & $0.1\,\mathrm{au} - 50\,\mathrm{au}$ \\
   $r_\text{out}$ & $100\,\mathrm{au} - 1000\,\mathrm{au}$ \\
   $M_\text{Dust}$ & $10^{-6}\,M_{\odot} - 10^{-3}\,M_{\odot}$ \\
   \hline
    \multicolumn{2}{|c|}{\bf{T Tauri star}} \\
   \hline
   $M_{\star}$ & $0.7\,M_{\odot}$ \\
   $R_{\star}$ & $2\,R_{\odot}$ \\
   $T_{\star}$ & $4000\,\mathrm{\kelvin}$\\
   \hline
    \multicolumn{2}{|c|}{\bf{Herbig Ae star}} \\
   \hline
   $M_{\star}$ & $2.5\,M_{\odot}$ \\
   $R_{\star}$ & $2.48\,R_{\odot}$ \\
   $T_{\star}$ & $9500\,\mathrm{\kelvin}$\\
   \hline
  \end{tabular}
  \caption{Model parameters}
  \label{tab:model}
 \end{table}
 
 If not specified otherwise, the results presented in the following are for a reference disk with an inner radius of
 $1\,\mathrm{au}$, a
 dust mass of $10^{-4}\,M_{\odot}$, $\beta=1.2$, $\alpha=2.1,$ and $h_\text{100}=10\,\mathrm{au}$. The inclination of
 the reference disk is $0^\circ$ (face-on), the wavelength is $9.5\,\mathrm{\mu\meter,}$ and the central star is a T Tauri star
 (see Tab.\,\ref{tab:model} for details).
 
 \subsection{Dust}
 \label{sec:dust}
 
 The dust model in our simulations consists of a homogeneous mixture of silicate and graphite adapted from 
 \cite{1984ApJ...285...89D}, \cite{1993ApJ...402..441L}, and \cite{2001ApJ...548..296W}.
 The dust mixture has an assumed density of $2.5\,\mathrm{g/cm^3}$ and consists of $62.5\,\mathrm{\%}$ silicate and $37.5\,\mathrm{\%}$ graphite.
 The one-third to two-third approximation by \cite{1993ApJ...414..632D} was applied to account for the anisotropic optical properties of graphite.
We applied a grain radius distribution following $n\sim a^{-3.5}$ with a lower grain radius of $5\,\mathrm{\nano\meter}$ and an upper grain radius of 
$250\,\mathrm{\nano\meter}$, which is consistent with the dust parameters found for the interstellar medium (ISM; \citealt{1977ApJ...217..425M}). 
Additionally, we performed simulations with an upper grain radius of $2.5\,\mathrm{\mu\meter}$ and $10\,\mathrm{\mu\meter}$ to account for
the presence of micron-sized grains (\citealt{2013A&A...549A.112M}). If not specified otherwise, the results presented in the following are for 
a reference grain size distribution with a maximum grain radius of $250\,\mathrm{\nano\meter}$.
 
 \section{Results}
 
We investigated the influence of the following parameters on the polarization state of the scattered light: 
the wavelength of the scattered radiation, the dust mass, and
 the inner disk radius. Afterward, we included the spatial dust distribution, disk inclination, and luminosity of the central star.
 Before assessing the influence of these parameters on the linear polarization degree (Sect.\,\ref{sec:polarisation}), we analyzed the 
 ratio of the thermal dust re-emission radiation to scattered stellar radiation (Sect.\,\ref{sec:ratio}).
 
 \subsection{Assessing the influence of the radiation source}
 \label{sec:ratio}
 

We assessed the relative influence of thermal re-emission radiation versus that of scattered
stellar radiation on the total flux.

\begin{figure}
 \centering
 \begin{Large}
\includegraphics[width=1.0\columnwidth]{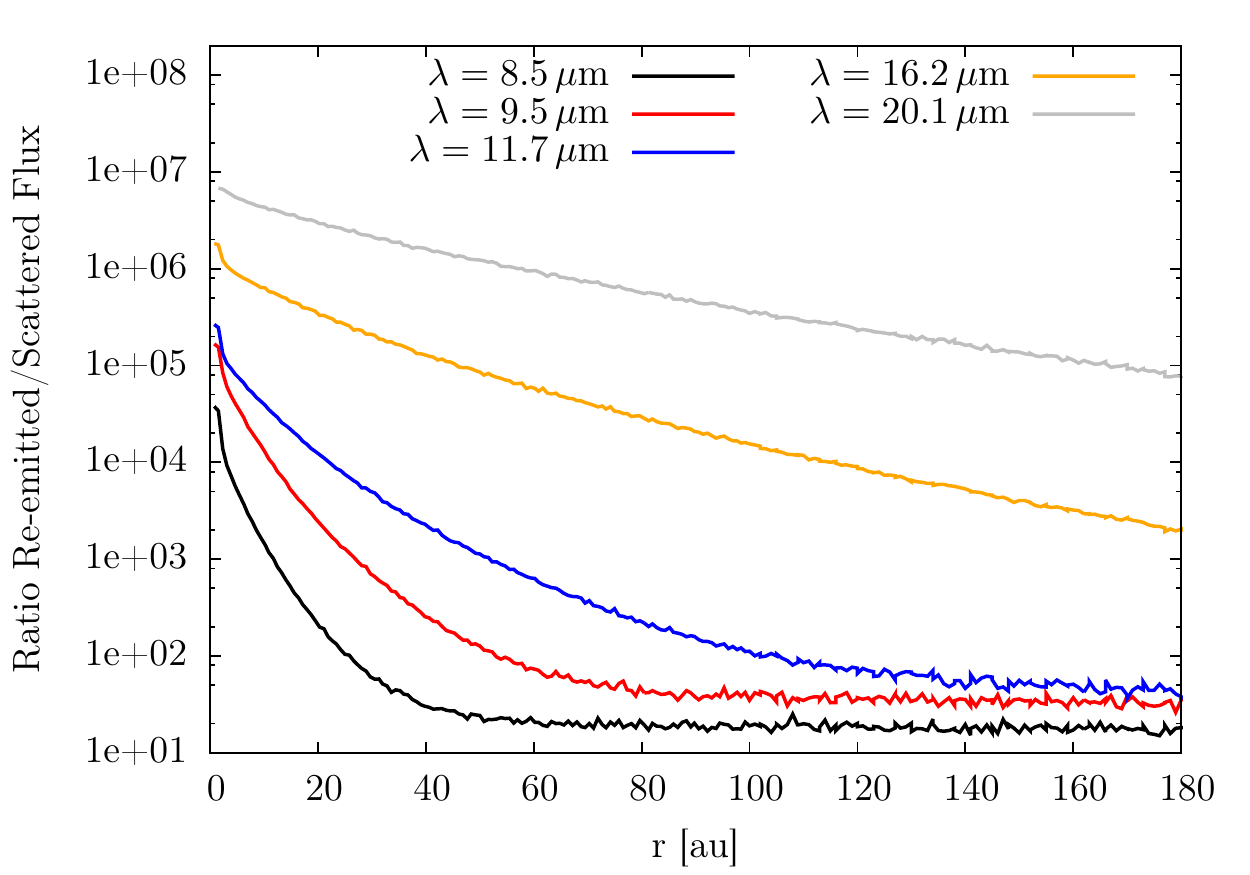}
\end{Large}
 \caption{Ratio of thermal re-emission radiation to scattered stellar radiation for
  wavelengths across the \textit{N} and \textit{Q} bands. For details, see Sect.\,\ref{sec:ratio}.}
 \label{img:ratio_1e-4_ro200_wl}
\end{figure}

\textbf{Wavelength:} The
 ratio
of both contributions across the \textit{N} and \textit{Q} bands
is shown in Fig.\,\ref{img:ratio_1e-4_ro200_wl}.
We find two important trends. First, close to the inner rim of the disk, the ratio reaches its maximum and decreases toward larger radial distances. This results from the temperature structure
 of the disk. The re-emitted flux decreases faster with
 increasing radial distance from the central star than the scattered stellar radiation does ($\sim r^{-2}$, geometrical attenuation). Second, because of the increasing contribution of the thermal re-emission radiation
 and the decreasing contribution of the stellar irradiation, the flux ratio increases with wavelength.

The general trends are the same if the central star is a Herbig Ae star. 
However, the ratio is increased because of the increased heating of the disk.

\textbf{Dust mass:} The flux ratio increases slightly with increasing mass, because the larger dust mass increases the thermal re-emission
radiation. However, the differences between the ratios are smaller than for a change in wavelength. 

\textbf{Outer disk radius:} The flux ratio increases slightly with increasing outer radius because the lower dust density
decreases the scattered stellar radiation. However, the differences between the ratios are even smaller than for a change in
dust mass.

\begin{figure}
 \centering
 \begin{Large}
 \includegraphics[width=1.0\columnwidth]{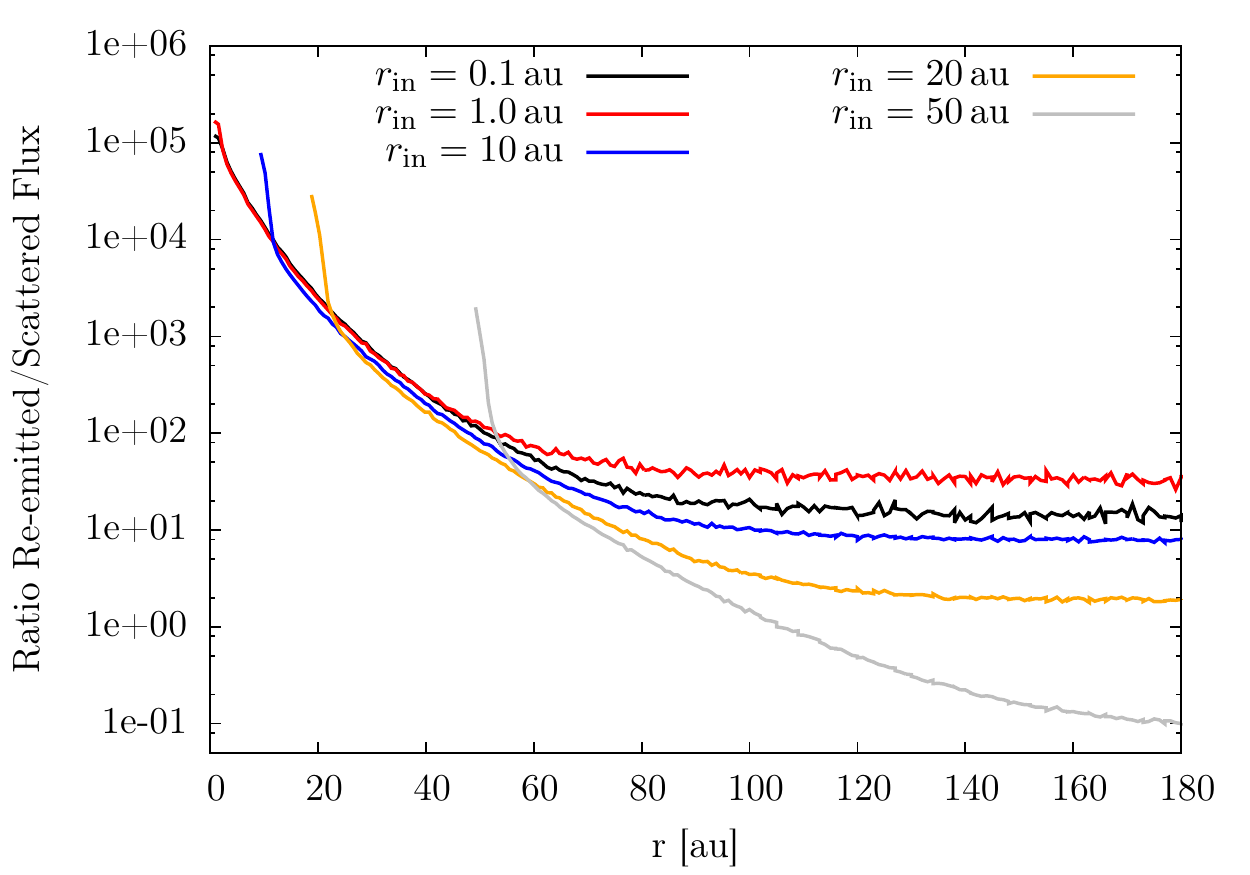}
 \end{Large}
 \caption{Ratio of thermal re-emission radiation to scattered stellar radiation for different inner radii.
 For details, see Sect.\,\ref{sec:ratio}.}
 \label{img:ratio_1e-4_ro200}
\end{figure}

\begin{figure}
  \centering
  \begin{Large}
  \includegraphics[width=1.0\columnwidth]{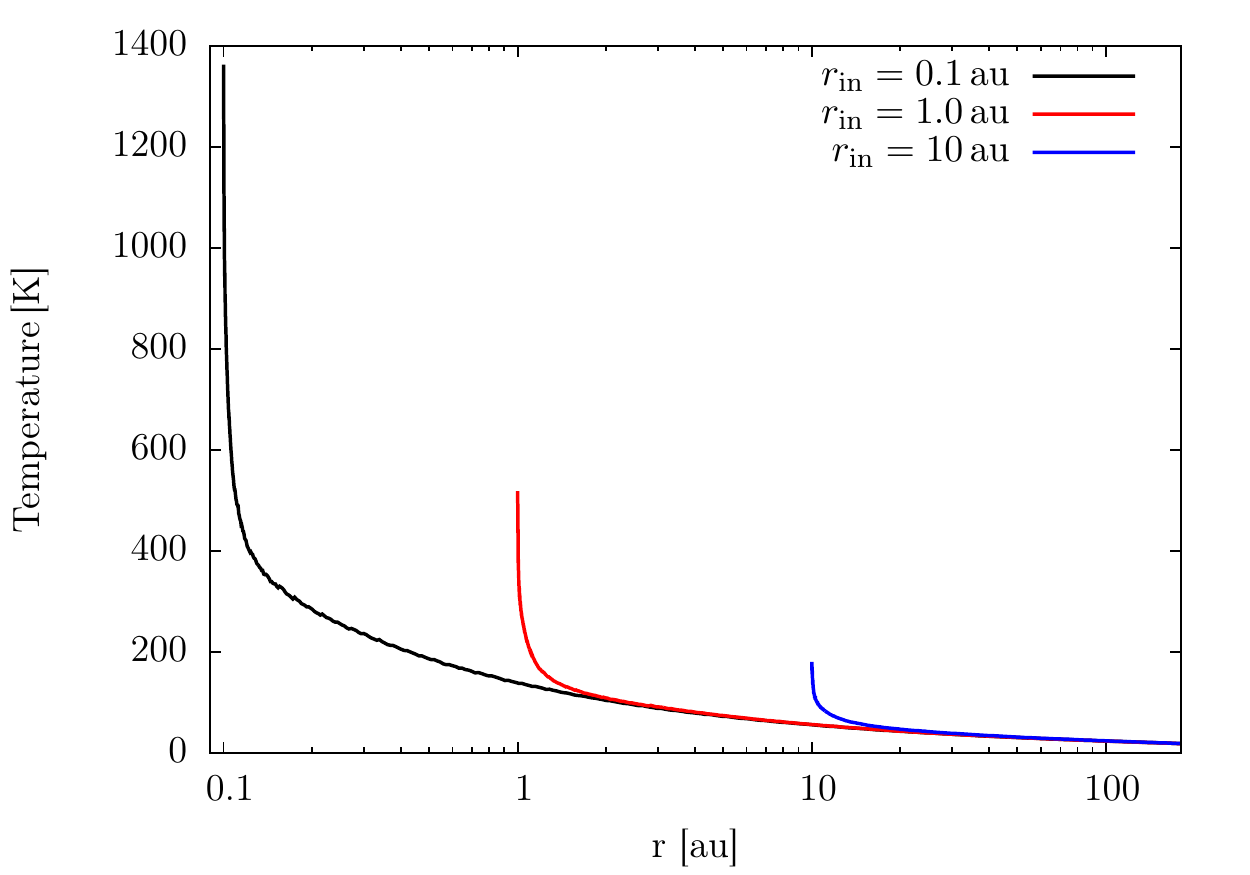}
  \end{Large}
  \caption{Comparison of the temperatures at the optical depth $\tau(9.5\,\mathrm{\mu\meter})=1$ surface as seen from above the disk, 
  perpendicular to the disk midplane, vs. radial distance for disks with different inner radii ($r_\text{in}$). 
  For details, see Sect.\,\ref{sec:ratio}.}
  \label{img:Innenradius_temp}
 \end{figure}

\textbf{Inner disk radius:} Shown in Fig.\,\ref{img:ratio_1e-4_ro200} is the ratio of 
thermal re-emission radiation to scattered stellar radiation for a fixed wavelength of $9.5\,\mathrm{\mu\meter}$
but different inner radii. 
As the temperature at the inner rim decreases with increasing inner disk radius, 
the contribution of the thermal re-emission radiation decreases 
while that of the stellar irradiation is hardly affected.
Interestingly, this trend is inverted for a disk with an inner radius of $0.1\,\mathrm{au}$.
That is because in this case, the location at which the temperature reaches $\sim 300\,\mathrm{\kelvin}$ moves
closer to the central star. In turn, the size of the effective mid-IR emitting region is reduced.
This is illustrated in Fig.\,\ref{img:Innenradius_temp} where the temperature of the optical depth $\tau(9.5\,\mathrm{\mu\meter})=1$ surface (as seen
from above the disk) is shown.

\begin{figure}
 \centering
 \begin{Large}
 \includegraphics[width=1.0\columnwidth]{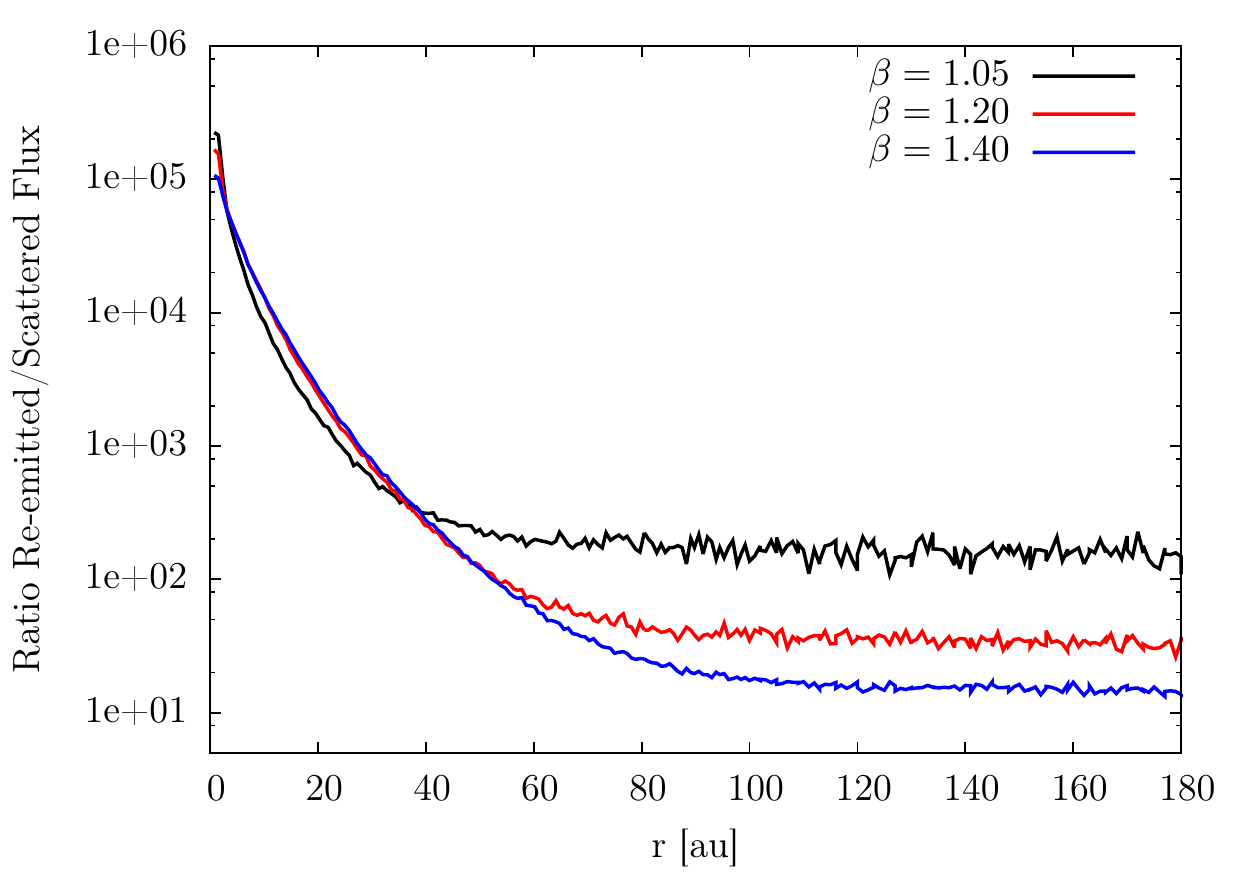}
 \end{Large}
 \caption{Ratio of thermal re-emission radiation to scattered stellar radiation for disks with
 different flaring exponents $\beta$.
 For details, see Sect.\,\ref{sec:ratio}.}
 \label{img:ratio_b}
\end{figure}

\textbf{Flaring exponent $\beta$:} A comparison of the flux ratios for different values of the flaring exponent $\beta$ is
shown in Fig.\,\ref{img:ratio_b}. For larger $\beta$, the larger increase of the scale height increases the efficiency
of light scattering and thus the intensity of scattered stellar radiation. The effective absorption of
the disk is increased as well, resulting in a more efficient dust heating. For small radial distances, the impact on
the dust heating dominates, while the impact on the scattering dominates further outside. 

\begin{figure}
 \centering
 \begin{Large}
 \includegraphics[width=1.0\columnwidth]{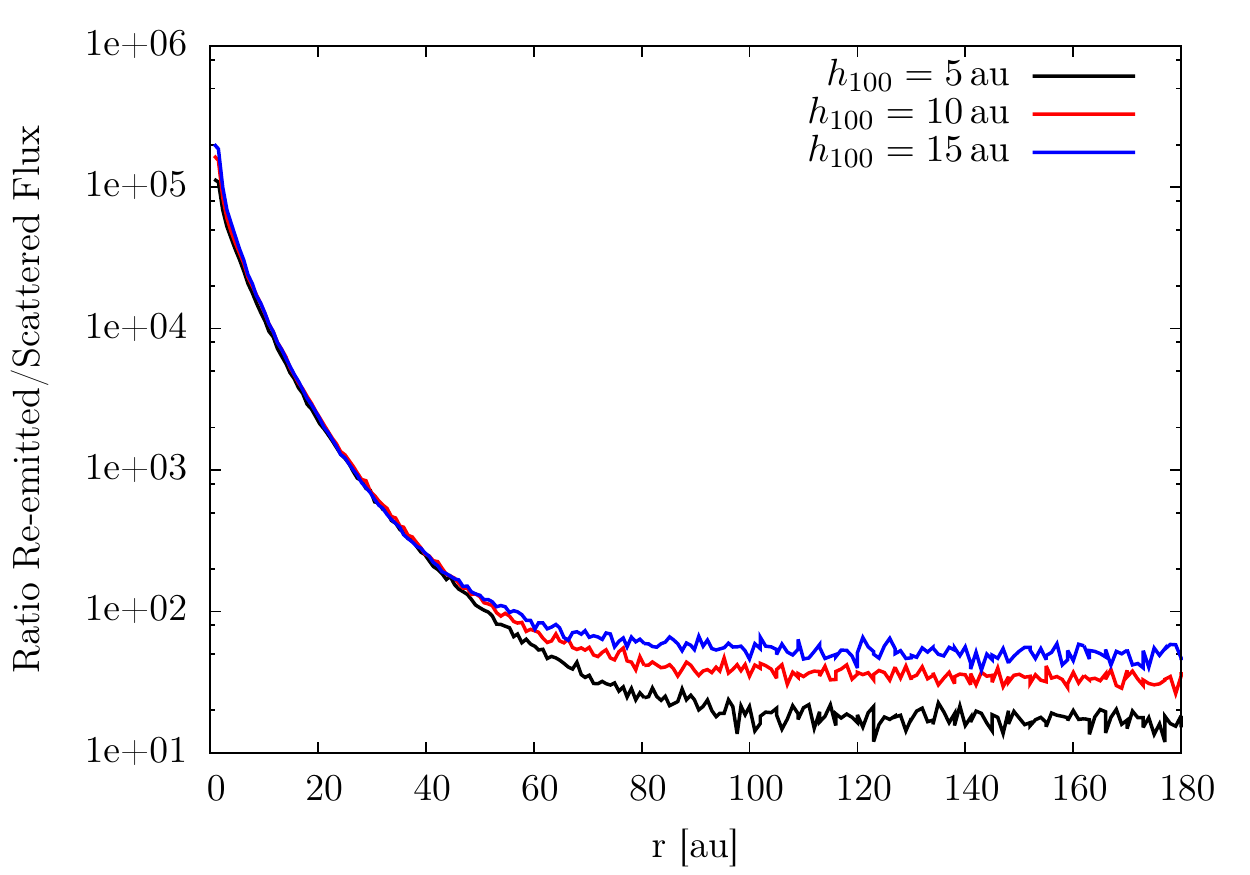}
 \end{Large}
 \caption{Ratio of thermal re-emission radiation to scattered stellar radiation for disks with
 different scale heights $h_\text{100}$.
 For details, see Sect.\,\ref{sec:ratio}.}
 \label{img:ratio_h}
\end{figure}

\textbf{Scale height $h_\text{100}$:} The flux ratio for different scale heights $h_\text{100}$ is shown in Fig.\,\ref{img:ratio_h}. Outside 
$\approx40\,\mathrm{au}$, the flux ratio significantly increases with increasing scale height, while the differences become minor further inside. 
A smaller scale height corresponds to a more compact disk in the vertical direction,
decreasing the width of the layer in which mid-IR radiation is produced.  
However, while the scattering of the stellar radiation is affected in a similar manner, 
the impact on the dust heating and re-emission dominates.

\textbf{Density distribution exponent $\alpha$:} Finally, we considered the exponent $\alpha$, which determines the radial decrease of the density distribution.
At steeper radial dust profiles the flux ratio increases. 
A more detailed description of this trend is omitted because the relative differences between the ratios are
smaller than for a change in $\beta$ or $h_\text{100}$.
 
 \textbf{Larger grains:} Extending the grain size distribution to larger radii ($2.5\,\mathrm{\mu\meter}$
 and $10\,\mathrm{\mu\meter}$)
 increases the net albedo (Fig.\,\ref{img:albedo}) of the dust and therefore the flux density of the scattered radiation.
 This decreases the ratio of thermal re-emission radiation to scattered stellar radiation,
 making the contribution of the scattered stellar radiation important at an inner radius of $\sim10\,\mathrm{au}$ already. However, we omitted a more detailed analysis because the trends found and discussed for small grains are not 
 altered qualitatively.
 
 In summary, the mid-IR thermal re-emission radiation of the dust is stronger
 than the scattered stellar radiation
 for disks with an inner radius up to $\sim20\,\mathrm{au}$ for grain size distributions with
 small maximum grain radii ($250\,\mathrm{\nano\meter}$) and up to $\sim10\,\mathrm{au}$ for 
 grain size distributions with larger maximum grain radii ($2.5\,\mathrm{\mu\meter}$, $10\,\mathrm{\mu\meter}$).
 Hence, changes of the thermal re-emission radiation potentially have
 a greater influence on the resulting linear polarization degree than modifications of the 
 scattered stellar radiation.
 However,
 for disks with larger inner radii (i.e., $r_\text{in}\gtrsim10\,\mathrm{au}$), the influence of the scattered stellar radiation 
 may become
 significant. This finding can be relevant for the analysis of polarization observations of
 transition disks, which often possess inner disk radii of a few dozen  astronomical units (e.g., \citealt{doi:10.1146/annurev-astro-081710-102548}).
 Therefore, both sources of radiation have to be taken into account when 
 analyzing observations of scattering polarization.
  
  \subsection{Scattered light polarization}
  \label{sec:polarisation}
  
  We investigated the dependence of the radial polarization profile on the radiation source (Sect.\,\ref{sec:quelle}),
  wavelength used (Sect.\,\ref{sec:vergleich_wl}), disk mass (Sect.\,\ref{sec:vergleich_mass}), and inner disk radius 
  (Sect.\,\ref{sec:vergleich_ri}). Afterward, we investigated the dependence of the radial polarization profile on the disk flaring,
  density distribution, and scale height in Sect.\,\ref{sec:param}, before we assessed the inclination dependence of the polarization pattern in
  Sect.\,\ref{sec:polmap} and the consequences of exchanging the central star with a Herbig Ae star in Sect.\,\ref{sec:vergleich_herbig}.
  
  \subsubsection{Source dependence of scattering polarization}
  \label{sec:quelle}
  
  \begin{figure}
 \centering
 \begin{Large}
 \includegraphics[width=1.0\columnwidth]{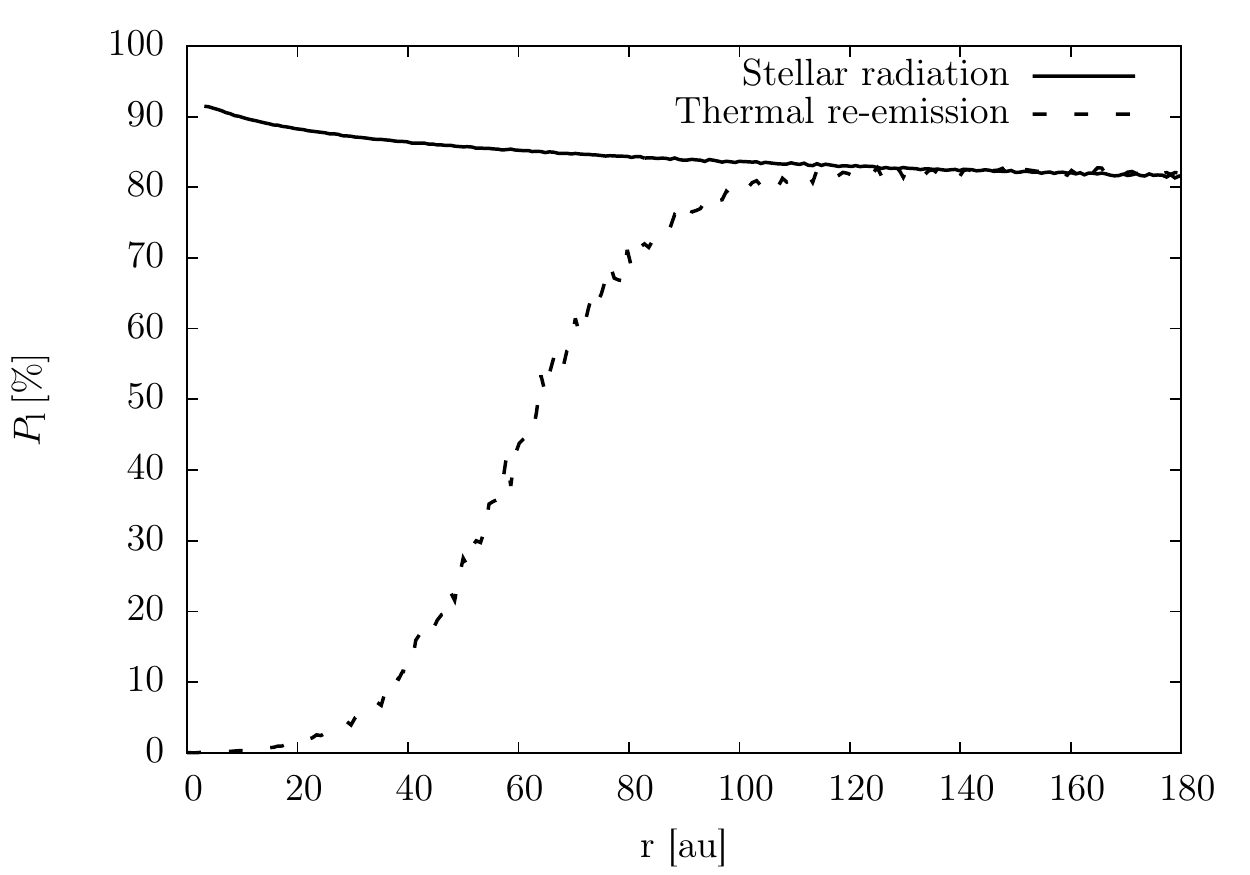}
 \end{Large}
 \caption{Comparison of the radial dependence of the linear polarization degree for scattered
 stellar radiation and scattered thermal re-emission radiation. 
 For details, see Sect\,\ref{sec:quelle}.}
 \label{img:source}
\end{figure}
  
  The individual contributions of scattered stellar radiation and scattered thermal re-emission
  radiation are shown in Fig.\,\ref{img:source}. For thermal re-emission radiation,
  the radiation originating at the inner part of the disk is almost unpolarized because its unscattered fraction dominates. Further outside, i.e., toward regions with lower dust temperatures, the contribution from the 
  direct thermal re-emission radiation decreases, while the contribution from the scattered radiation decreases less 
  rapidly. Consequently, the linear polarization degree of the thermal re-emission radiation increases with increasing
  distance from the central star. 
  
  For stellar radiation, there is no unscattered contribution besides the direct photospheric emission.
  The radiation is significantly polarized, even in the inner part
  of the disk. The linear polarization degree decreases slightly with increasing radial distance 
  because the scattering angle increasingly deviates from $90^\circ$ owing to the disk flaring.
  Outside a radial distance of $\approx 160\,\mathrm{au}$, where the unscattered
  thermal re-emission radiation is negligible, the linear polarization degree for both radiation sources becomes similar.
  When calculating the net polarization, taking both radiation sources (disk, star) into account, the 
  relative contribution of both sources to the net flux (discussed in Sect.\,\ref{sec:ratio}) becomes important.
  
  \begin{figure}
 \centering
 \begin{Large}
 \includegraphics[width=1.0\columnwidth]{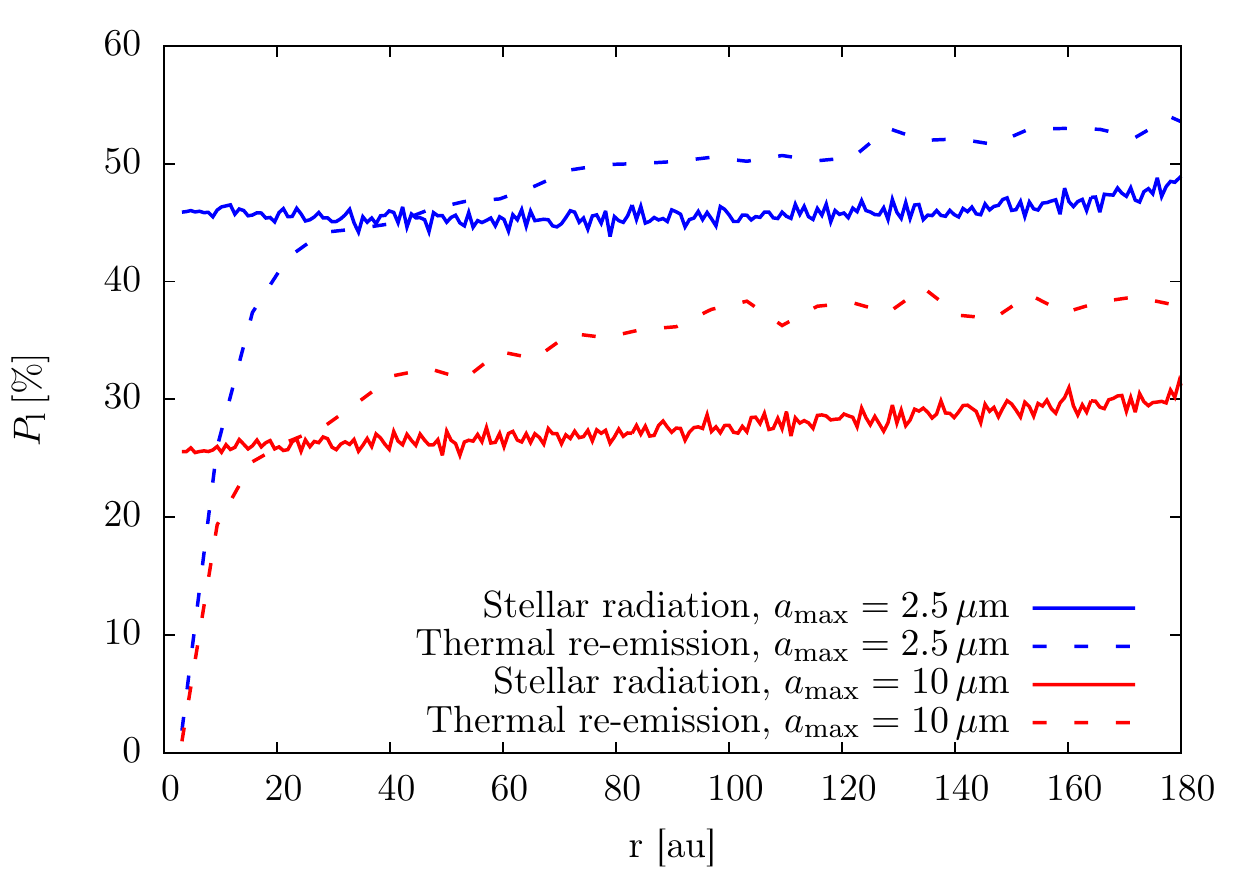}
 \end{Large}
 \caption{Radial polarization profiles at $9.5\,\mathrm{\mu\meter}$ for the case of grain size
 distributions with larger maximum grain radii ($2.5\,\mathrm{\mu\meter}$, $10\,\mathrm{\mu\meter}$).
 For details, see Sect\,\ref{sec:quelle}.}
 \label{img:source_large}
\end{figure}
  
  \begin{figure}
  \centering
 \begin{Large}
   \begin{subfigure}[b]{0.7\columnwidth}
   \includegraphics[width=1.0\columnwidth]{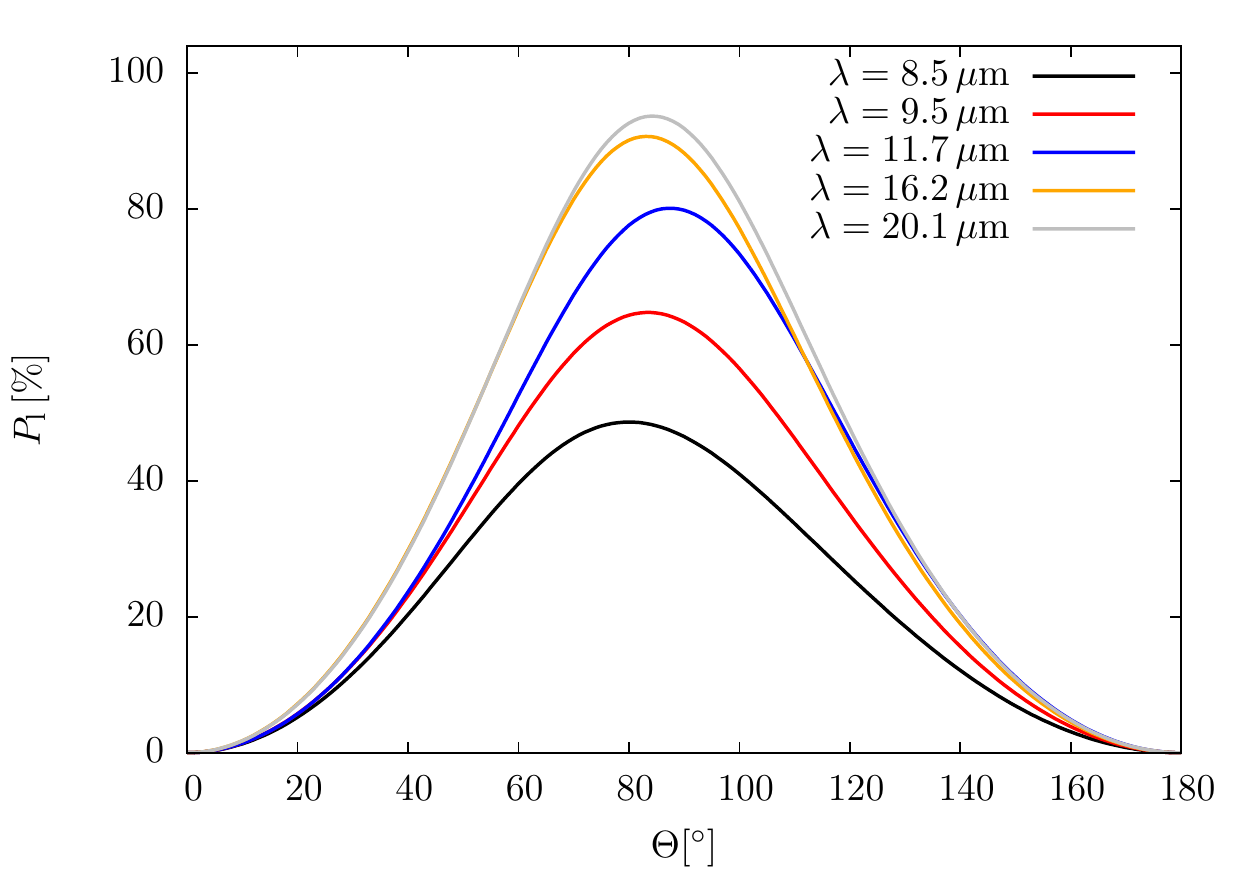}
   \end{subfigure}
   \begin{subfigure}[b]{0.7\columnwidth}
   \includegraphics[width=1.0\textwidth]{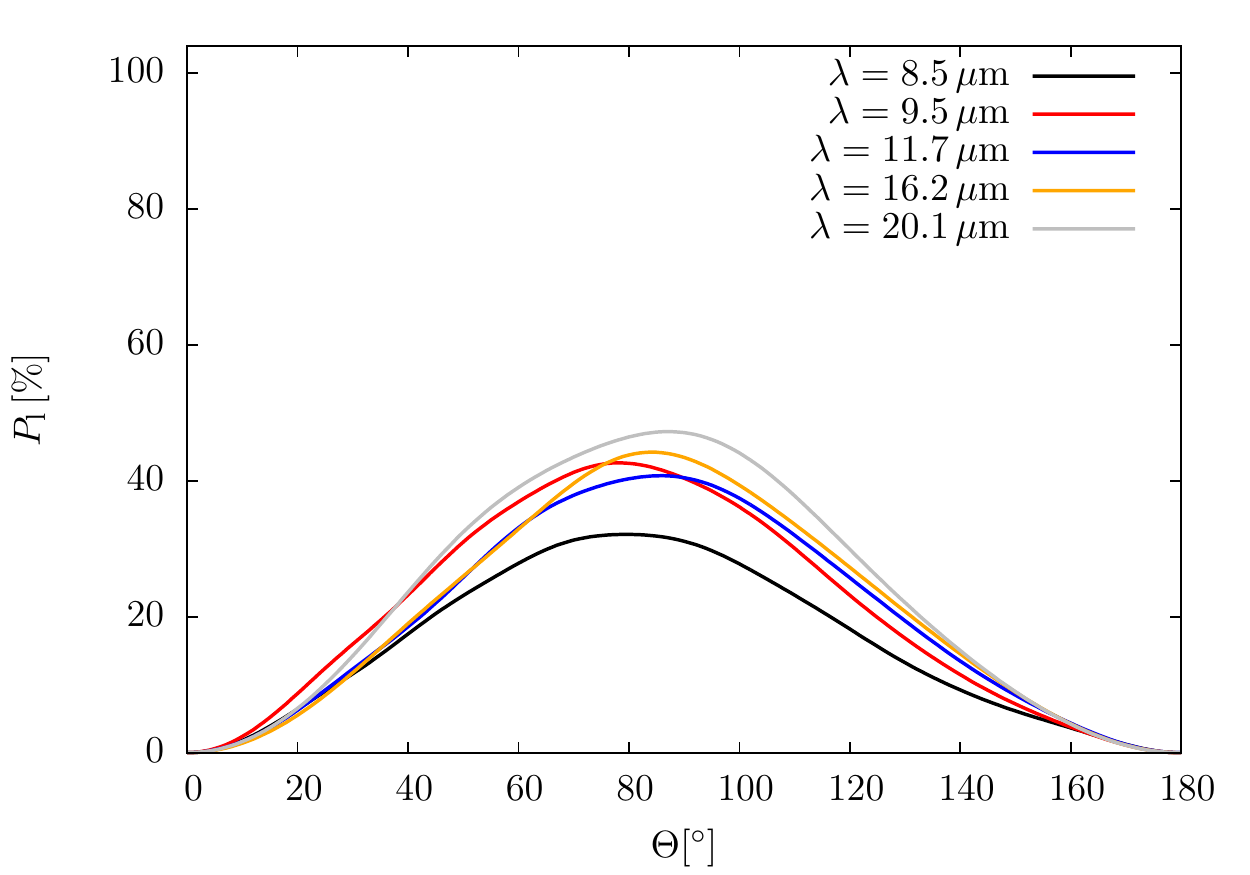}
   \end{subfigure}
    \end{Large}
    \caption{Comparison of the expected linear polarization degree as a function of the scattering angle $\Theta$
    for grain size distributions
    with a maximum grain radius of $a_\text{max}=2.5\,\mathrm{\mu\meter}$ (top)
    and $a_\text{max}=10\,\mathrm{\mu\meter}$ (bottom), for various wavelengths. For details, see Sect.\,\ref{sec:quelle} and \ref{sec:vergleich_wl}.}
    \label{img:P_expected}
  \end{figure}
  
  In the case of grain size distributions with a larger maximum grain radius, we find the following main differences to the above scenario
  based on ISM-like grain sizes (see Fig.\,\ref{img:source_large}):

   The overall polarization degree is lower. This is because the expected polarization degree for single scattering
   is lower for larger than for smaller grains (see Fig.\,\ref{img:P_expected}).
   
\begin{figure}
 \centering
 \begin{Large}
 \includegraphics[width=1.0\columnwidth]{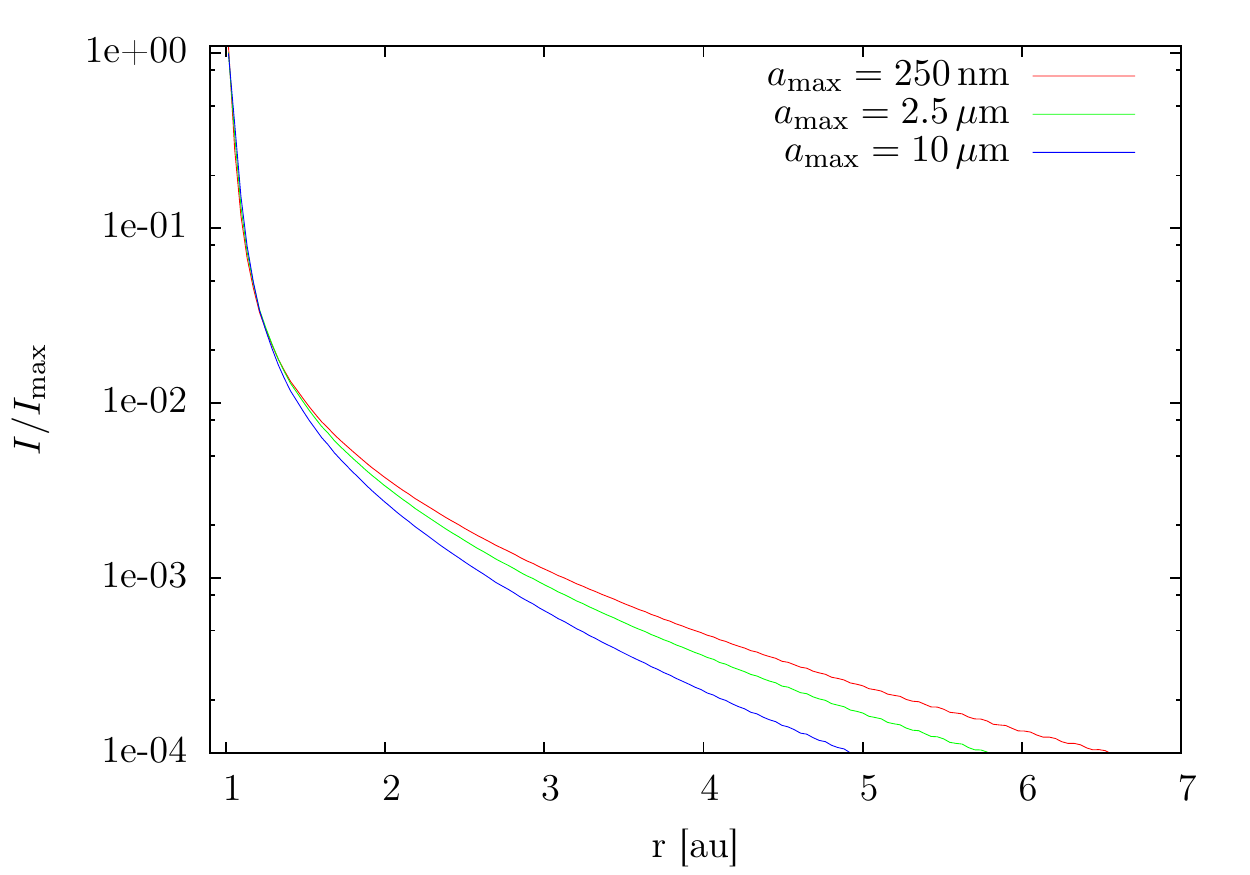}
 \end{Large}
 \caption{Comparison of the radial extent of the effective mid-IR emitting region (reference wavelength: $9.5\,\mathrm{\mu\meter}$).
 For details, see Sect\,\ref{sec:quelle}.}
 \label{img:ungestreut}
\end{figure}

   The polarization degree of scattered thermal re-emission radiation increases faster with radial distance than for a grain
   size distribution with a smaller maximum grain radius.
   This is because of the higher albedo of the grain size distributions with a larger maximum grain radius
   (see Fig.\,\ref{img:albedo}), resulting in an increased contribution of scattered and therefore polarized radiation.
   Additionally, the area of effective mid-IR radiation, see Fig.\,\ref{img:ungestreut}, is smaller
   because of the steeper decrease of the temperature with radial distance.
   Consequently, the mid-IR bright region in which the direct (unscattered) thermal re-emission radiation
   dominates the net flux (and thus the net polarization) is confined to a smaller region in the case of
   grain size distributions with a larger maximum grain radius.
   
\begin{figure}
 \centering
 \begin{Large}
 \includegraphics[width=1.0\columnwidth]{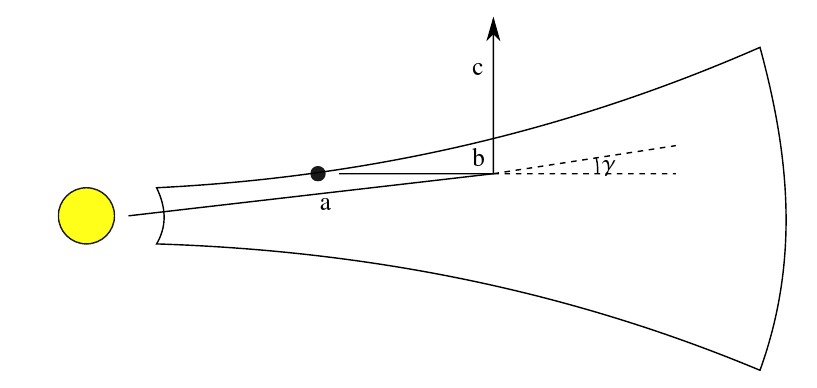}
 \end{Large}
 \caption{Comparison of the path stellar radiation (a) and radiation originating from thermal re-emission (b) take 
 toward the point of scattering
 and the path they take after scattering (c). The path the radiation would take without scattering is
 illustrated with a dashed line. The difference of the scattering angles for 
 stellar radiation and scattered thermal re-emission radiation is denoted as $\gamma$.
 For details, see 
 Sect.\,\ref{sec:quelle}.}
 \label{img:drawing2}
\end{figure}

\begin{figure}
 \centering
 \begin{Large}
 \includegraphics[width=1.0\columnwidth]{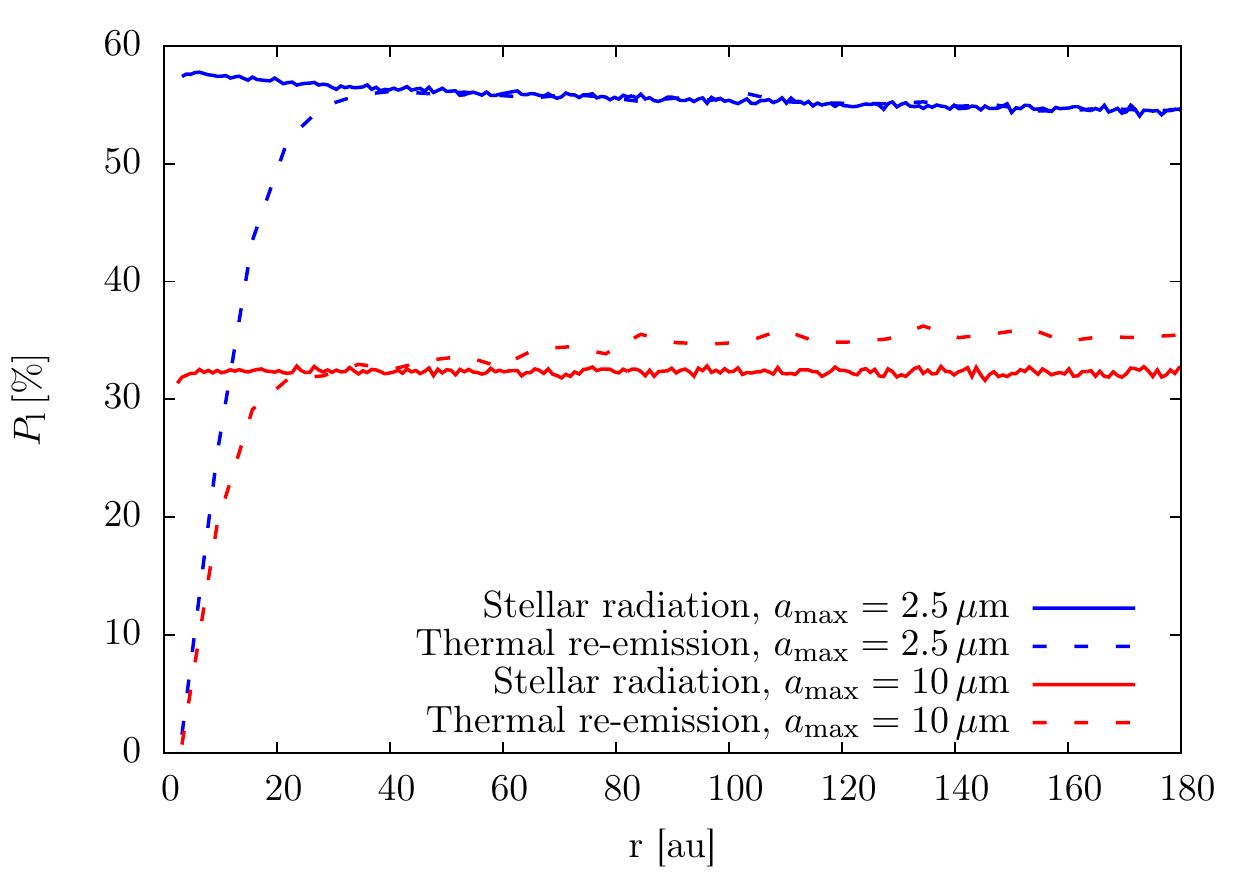}
 \end{Large}
 \caption{Radial polarization profiles at $9.5\,\mathrm{\mu\meter}$ for the case of grain size distributions with larger maximum grain radii
 ($2.5\,\mathrm{\mu\meter}$, $10\,\mathrm{\mu\meter}$). For illustration purposes (see Sect.\,\ref{sec:quelle}), only 
 the fraction of the radiation that has been scattered
 at most once is considered, i.e., multiply scattered radiation is not considered.}
 \label{img:0int}
\end{figure}

   The radial polarization profiles for scattered stellar radiation and scattered thermal re-emission radiation do not 
   converge to the same value. There are three reasons for this: 

First, the stellar radiation has to pass through the optically thick inner rim
   of the disk, 
   while the thermal re-emission radiation originates mainly from the surface of the disk (see Fig.\,\ref{img:drawing2}).
   Therefore, the depolarization by 
   multiple scattering is stronger for scattered stellar radiation than for scattered thermal re-emission radiation.
   For illustration of this effect, Fig.\,\ref{img:0int} shows that neglecting multiple scattering
   significantly reduces the difference between scattered stellar radiation
   and scattered thermal re-emission radiation.
   This depolarization
   effect is less important in the case of
   grain size distributions with small maximum grain radii because the albedo of these is
   so small that multiple scattering can be neglected (see Fig.\,\ref{img:albedo}).\\   
   Second, the scattering angles for scattered stellar radiation and scattered 
   thermal re-emission radiation are different (see Fig.\,\ref{img:drawing2}). 
   Although this applies for grain size distributions with small maximum grain radii as well, it only has a significant impact on the 
   net polarization for grain size distributions with larger maximum grain radii because of the increased contribution of scattered radiation.\\   
   Third, the effective radiation source for mid-IR thermal re-emission radiation is more compact for grain size distributions with larger maximum grain radii
   (see Fig.\,\ref{img:ungestreut}), increasing the anisotropy of the overall radiation field and thus the linear polarization degree.
   This effect is marginal for the large disks we consider in this work, but it can gain significance 
   in smaller disks.

  \subsubsection{Wavelength dependence of scattering polarization}
  \label{sec:vergleich_wl}
  
\begin{figure}
 \centering
 \begin{Large}
 \includegraphics[width=1.0\columnwidth]{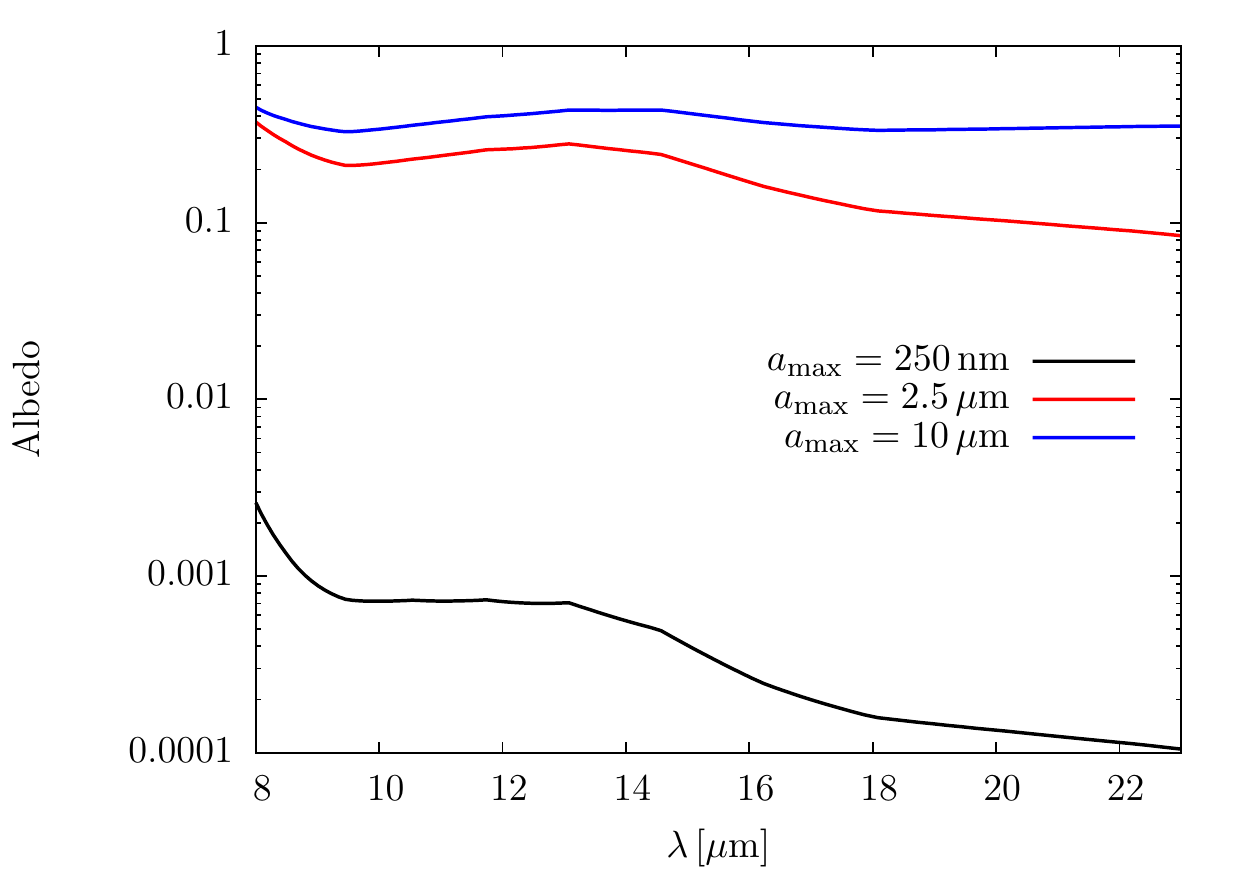}
 \end{Large}
 \caption{Wavelength-dependence of the dust albedo in the considered wavelength range.
 For details, see Sect.\,\ref{sec:vergleich_wl}.}
 \label{img:albedo}
\end{figure}

  The resulting linear polarization degree of scattering polarization is determined by the polarization after single scattering, the albedo and the 
  ratio of scattered radiation versus unscattered, i.e., unpolarized radiation.
  The wavelength dependence of the single scattering polarization is negligible because the shortest wavelength considered is
  much larger than the maximum grain radius.
  The scattering angle is expected to get closer to $90^\circ$ 
  because the height of the optical depth $\tau=1$-surface decreases with increasing wavelength.
  This would increase the linear polarization degree. However, the albedo (Fig.\,\ref{img:albedo})
  decreases, thereby decreasing the flux density of the scattered and thus polarized radiation. In addition,
  the flux density of the thermal re-emission radiation 
  increases with wavelength, thereby increasing the contribution of the unscattered radiation.

\begin{figure}
  \centering
\begin{Large}
  \includegraphics[width=1.0\columnwidth]{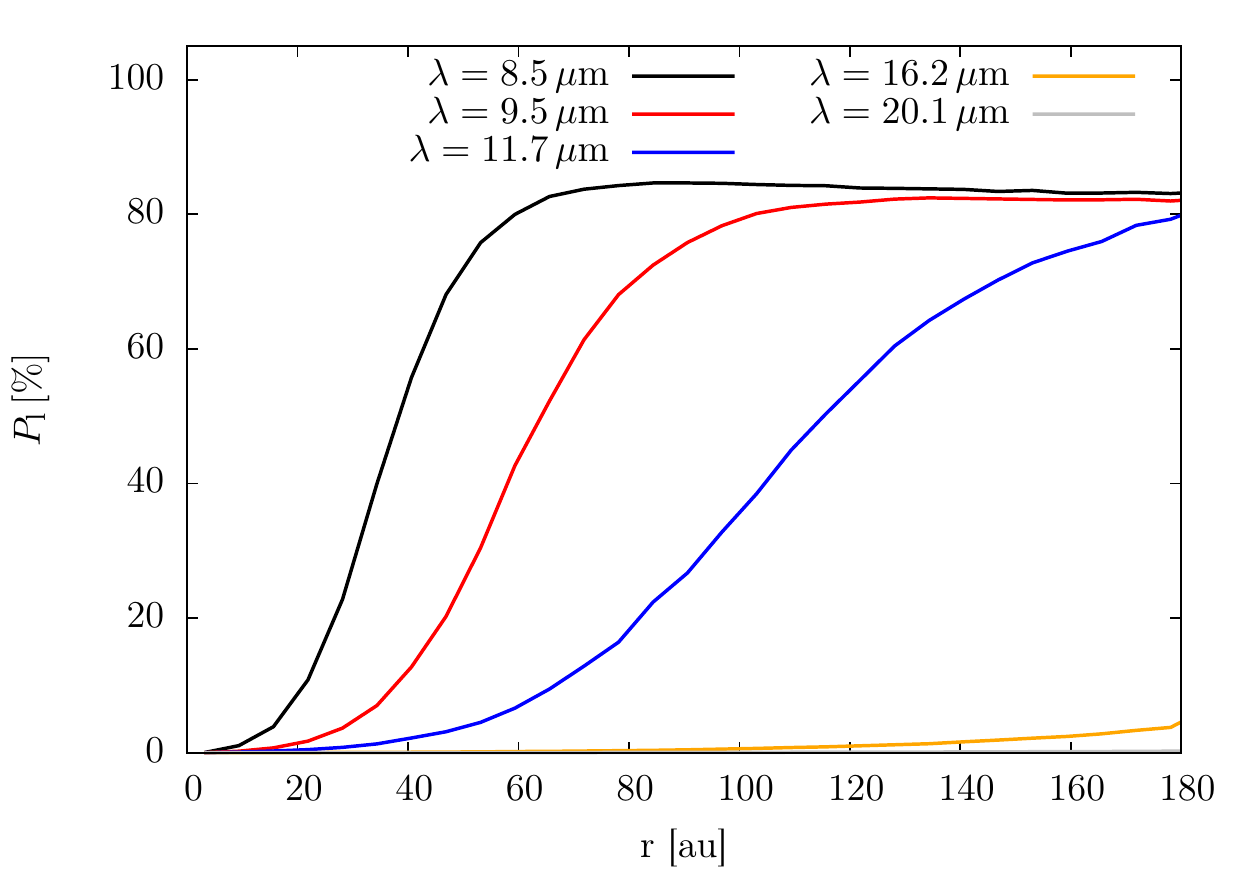}
  \end{Large}
\caption{Radial polarization profiles at
wavelengths across the \textit{N} and \textit{Q} bands. 
For details, see Sect.\,\ref{sec:vergleich_wl}.}
\label{img:vergleich_wl}
 \end{figure}

  To test which mechanism dominates, the radial polarization profiles for multiple
  wavelengths through the \textit{N} and \textit{Q} bands are computed and the results are shown in Fig.\,\ref{img:vergleich_wl}.
  The linear polarization degree decreases with increasing wavelength, indicating that the change in the ratio of scattered versus
  unscattered radiation dominates that of the scattering angle.
  In addition, the linear polarization degree increases with radial distance, similar to the radial dependence of
  the linear polarization degree for pure thermal re-emission radiation (Sect.\,\ref{sec:quelle}).
  
  \begin{figure}
  \centering
 \begin{Large}
   \begin{subfigure}[b]{0.7\columnwidth}
   \includegraphics[width=1.0\textwidth]{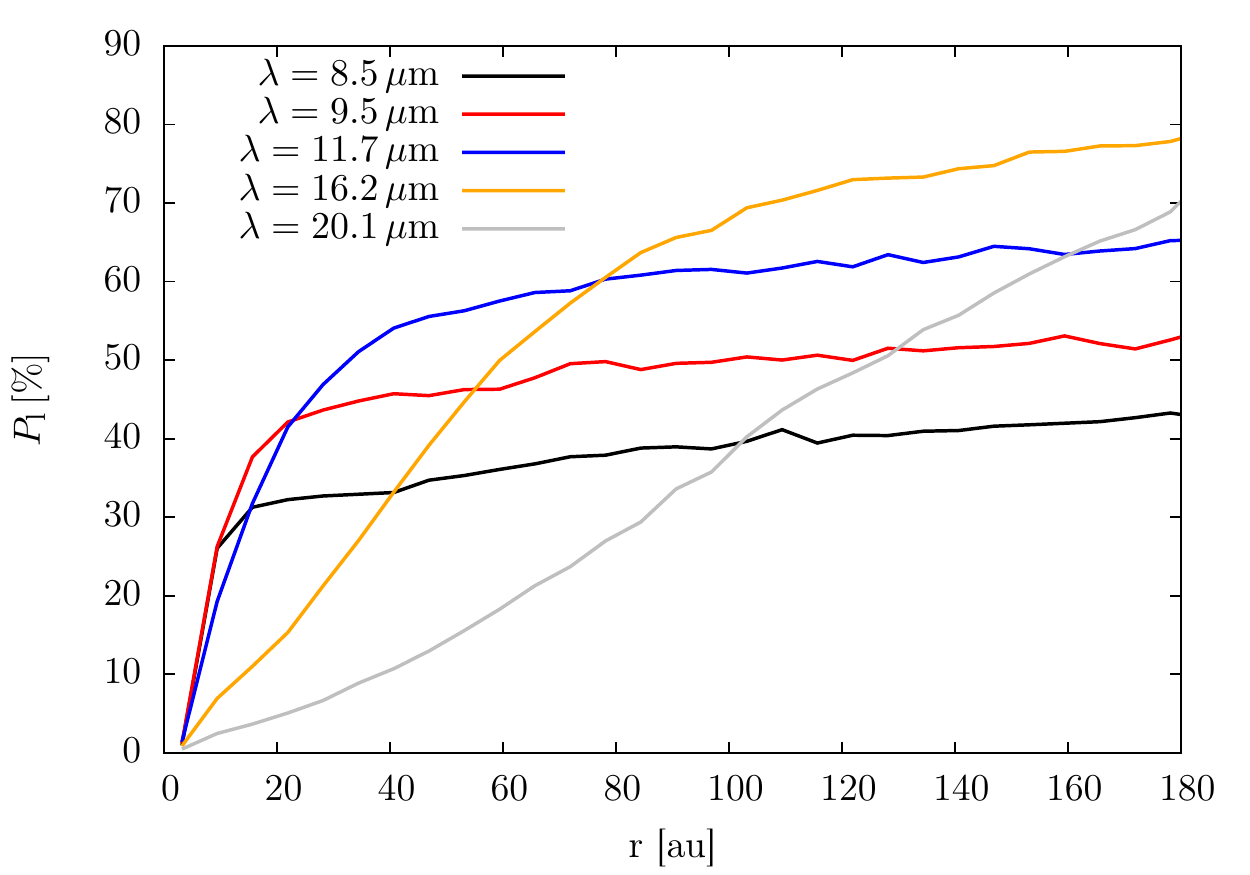}
   \end{subfigure}
   \begin{subfigure}[b]{0.7\columnwidth}
   \includegraphics[width=1.0\textwidth]{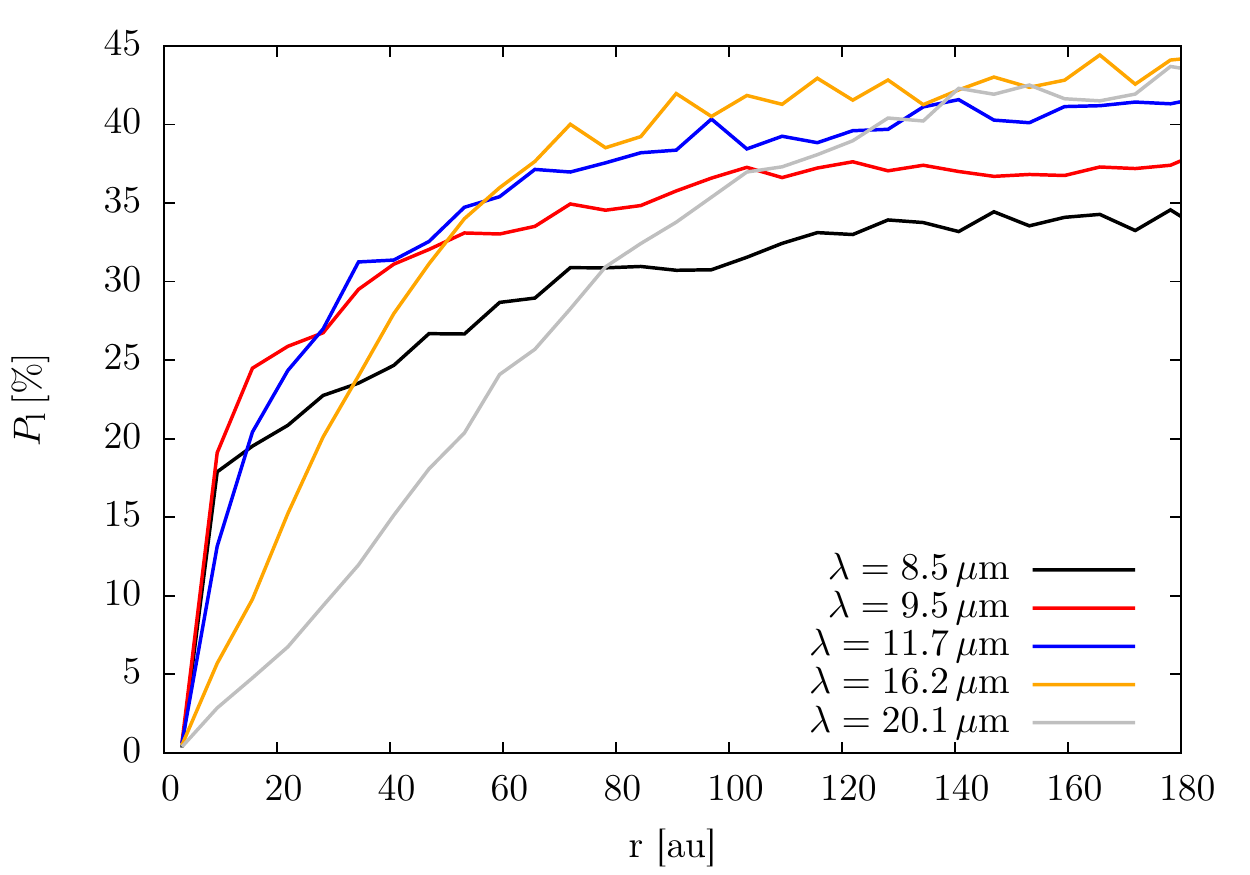}
   \end{subfigure}
    \end{Large}
    \caption{Radial polarization profiles for the case of grain size distributions with larger maximum grain radii
    (top: $2.5\,\mathrm{\mu\meter}$, bottom: $10\,\mathrm{\mu\meter}$) at wavelengths across the \textit{N} and \textit{Q} bands.
For details, see Sect.\,\ref{sec:vergleich_wl}.}
    \label{img:vergleich_wl_large}
  \end{figure}
  The radial polarization profiles for different wavelengths for the grain size distributions with 
  larger maximum grain radii ($2.5\,\mathrm{\mu\meter}$, $10\,\mathrm{\mu\meter}$) are shown
  in Fig.\,\ref{img:vergleich_wl_large}. The differences between these profiles and the profile for grain size distributions
  with small maximum grain radii ($250\,\mathrm{\nano\meter}$; Fig.\,\ref{img:vergleich_wl})
  can be explained as follows:
  For larger maximum grain radii, the albedo (Fig.\,\ref{img:albedo}) increases. Therefore, the flux density of the 
  scattered radiation is significantly increased. However, the expected linear polarization degree (Fig.\,\ref{img:P_expected})
  now depends on the wavelength. Therefore, the profiles for different wavelengths now converge to different values
  of the linear polarization degree. In addition, the depolarization due to multiple scattering needs to be taken into account.
  
  \subsubsection{Dependence of scattered light polarization on disk mass and outer radius}
  \label{sec:vergleich_mass}
  
  A lower disk mass results in a decreased optical depth of the disk, decreasing the height of the
  optical depth $\tau(9.5\,\mathrm{\mu\meter})=1$ surface (as seen from above the disk).
  Therefore, the scattering angle is expected to be closer to $90^\circ$, increasing the linear polarization degree
  at larger radial distances. Simultaneously, the lower optical depth enables the light from the star to penetrate 
  deeper into the disk, increasing the efficiency of dust heating. This increases the 
  contribution of unscattered radiation and therefore decreases the linear polarization degree
  in the innermost disk regions.
  
  \begin{figure}
   \centering
   \begin{Large}
   \includegraphics[width=1.0\columnwidth]{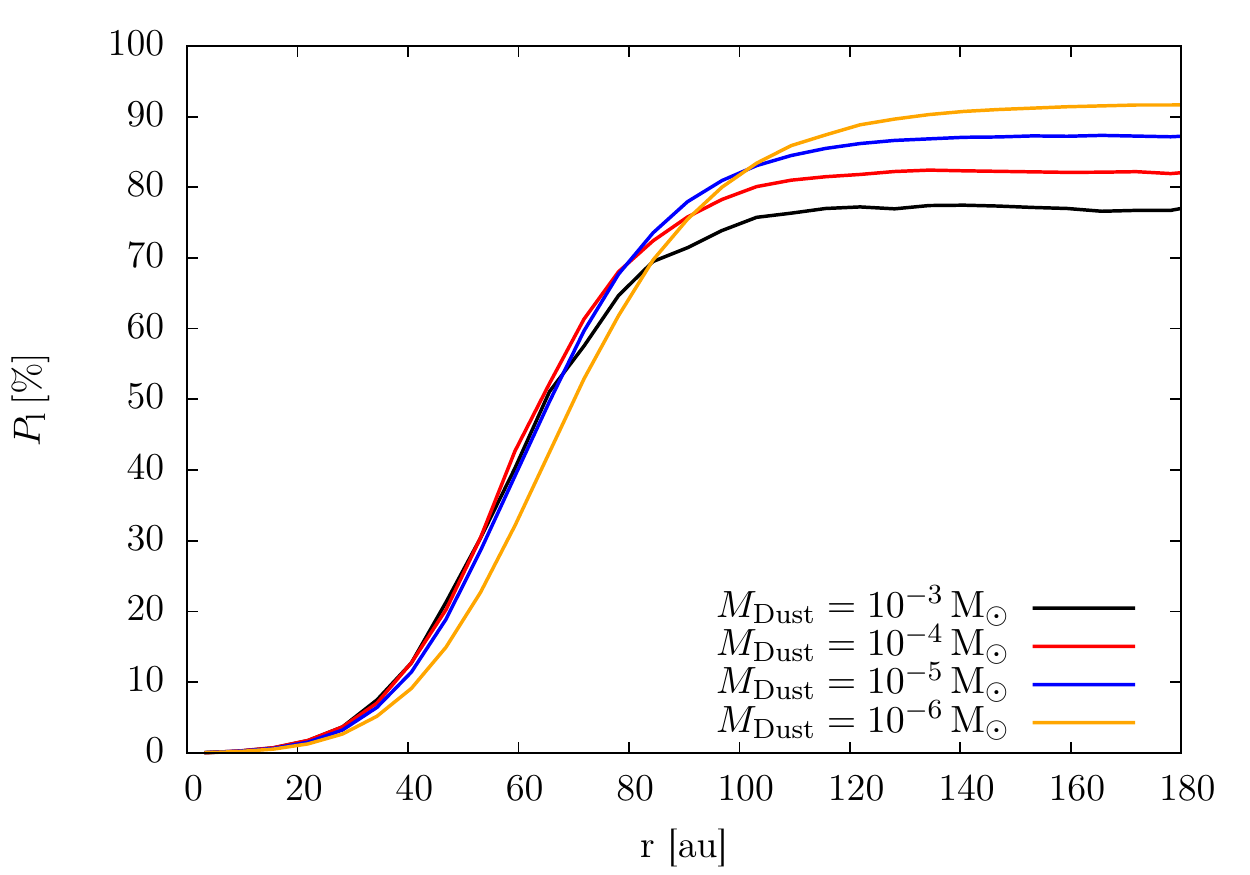}
   \end{Large}
   \caption{Radial polarization profiles for various disk masses.
   For details, see Sect.\,\ref{sec:vergleich_mass}.}
   \label{img:vergleich_mass}
  \end{figure}

  The radial polarization profiles for disks with various masses are shown in Fig.\,\ref{img:vergleich_mass}.
  As expected, the linear polarization degree for disks with lower masses is decreased at small radial distances,
  but increased further outside.
  
  For grain size distributions with medium ($2.5\,\mathrm{\mu\meter}$) maximum grain radii, the trends are similar. However, 
  the differences are reduced because the degree of the linear polarization (resulting from single scattering)
  is not a simple symmetric function of the scattering angle (Fig.\,\ref{img:P_expected}).
  More specifically, the maximum polarization occurs for a scattering angle slightly below $90^\circ$.
  For the largest maximum grain radii ($10\,\mathrm{\mu\meter}$), the polarization degree increases with disk mass for all radial distances
  because the maximum polarization occurs for a smaller scattering angle than for grains with smaller maximum grain radii.
  
  Increasing the outer radius decreases the optical depth as well. However, the differences in the optical depth are smaller
  than for a change of the disk mass, resulting in a minor
   change of the radial polarization profile with outer radius.
  
 \subsubsection{Dependence of scattered light polarization on inner disk radius}
 \label{sec:vergleich_ri}
 The temperature at the inner rim (Fig.\,\ref{img:Innenradius_temp}) decreases with increasing inner radius.
 This results in a decrease of the overall flux density and thus the contribution by self-scattered radiation.
 
  \begin{figure}
  \centering
\begin{Large}
  \includegraphics[width=1.0\columnwidth]{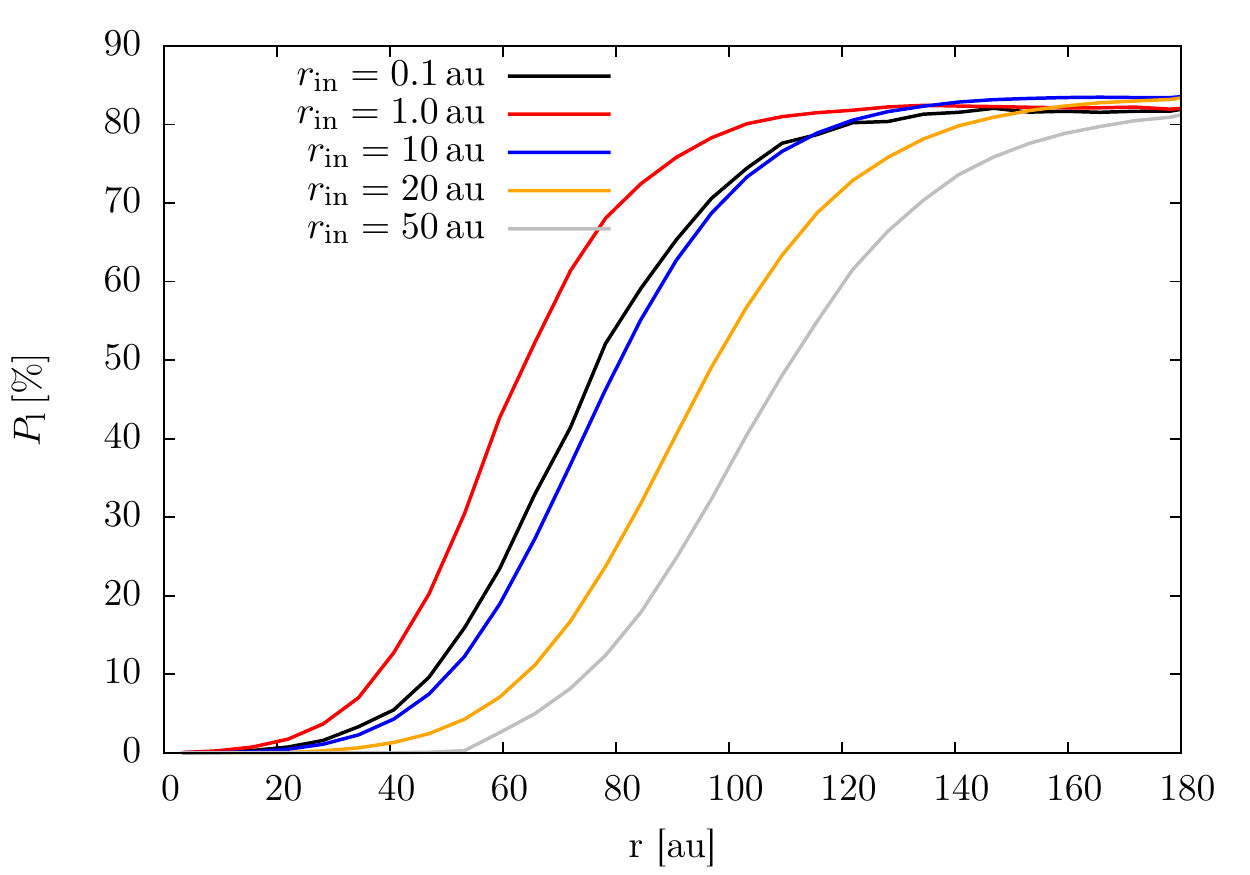}
  \end{Large}
\caption{Radial polarization profiles for various
inner disk radii. 
For details, see Sect.\,\ref{sec:vergleich_ri}.}
\label{img:vergleich_ri}
 \end{figure}
 
 In Fig.\,\ref{img:vergleich_ri} we show the radial polarization profiles for disks with various inner radii.
 The increase of the linear polarization degree is shifted outside for larger inner radii
 because the inner rim of the disk is shifted further outside.
 Interestingly, this trend is reversed for an inner radius of $0.1\,\mathrm{au}$.
 This is because for the disk with an inner radius of $0.1\,\mathrm{au}$, the radial distance where
 the temperature becomes $\sim 300\,\mathrm{\kelvin}$ is located closer to the central star than for the disk with an inner radius of
 $1\,\mathrm{au}$. As a result, the size of the effective mid-IR emitting region is reduced. 
 Grain size distributions with larger maximum grain radii show similar trends. However, the maximum polarization degree is altered
 as well for different inner radii.

In summary, the linear polarization degree increases at the smallest radial distances if
the size of the effective mid-IR emitting region is maximized.
 Both, smaller and larger corresponding inner radii shift this increase of the linear polarization degree to larger radial distances.

\subsubsection{Influence of disk flaring, density distribution, and scale height}
\label{sec:param}

In the preceding sections, we assumed a flaring exponent of $\beta=1.2$, radial density profile
described by $\alpha=2.1,$ and scale height of $h_\text{100}=10\,\mathrm{au}$.
In the following, we investigate the influence of these parameters
on the radial polarization profile.

\begin{figure}
 \centering
 \begin{Large}
 \includegraphics[width=1.0\columnwidth]{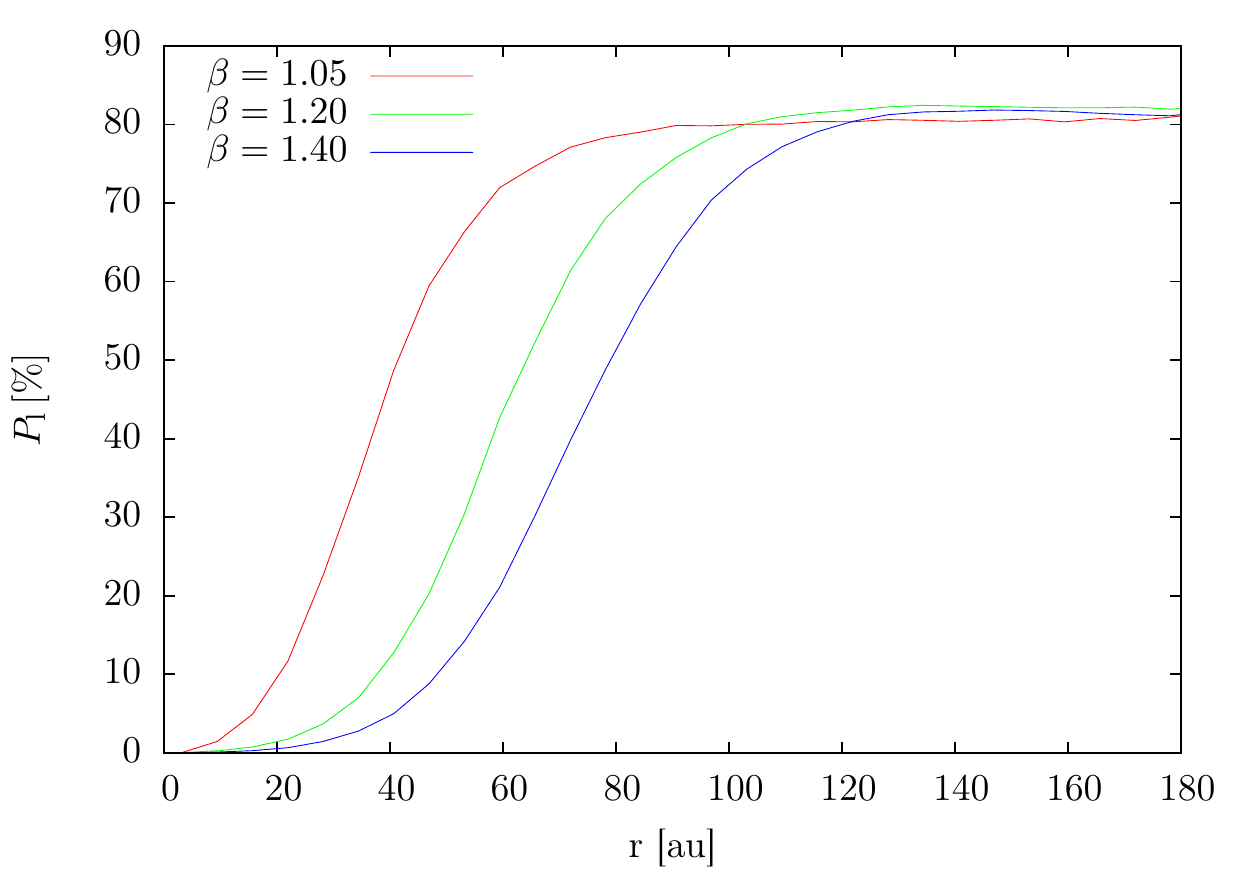}
 \end{Large}
 \caption{Comparison of the radial polarization profiles for various flaring exponents.
  For details, see Sect. \ref{sec:param}.}
 \label{img:vergleich_b}
\end{figure}

\textbf{Flaring exponent $\beta$:} The radial dependence of the linear polarization degree for different flaring exponents $\beta$ is shown in
Fig.\,\ref{img:vergleich_b}. 
The relative difference in the linear polarization degree is largest for intermediate radial distances, i.e., between $\approx40\,\mathrm{au}$
and $\approx 80\,\mathrm{au}$. There, the linear polarization degree decreases with increasing $\beta$, i.e., increasing disk flaring.
 The greater increase of the 
gradient of the scale height significantly increases the effective absorption of the disk, resulting in a more efficient
dust heating and therefore larger contribution by unscattered thermal re-emission radiation. 
If a grain size distribution with larger maximum grain radii is used, the trend is similar, but the differences are largest for small radial distances,
i.e., inside $\sim 20\,\mathrm{au}$.

\begin{figure}
 \centering
 \begin{Large}
 \includegraphics[width=1.0\columnwidth]{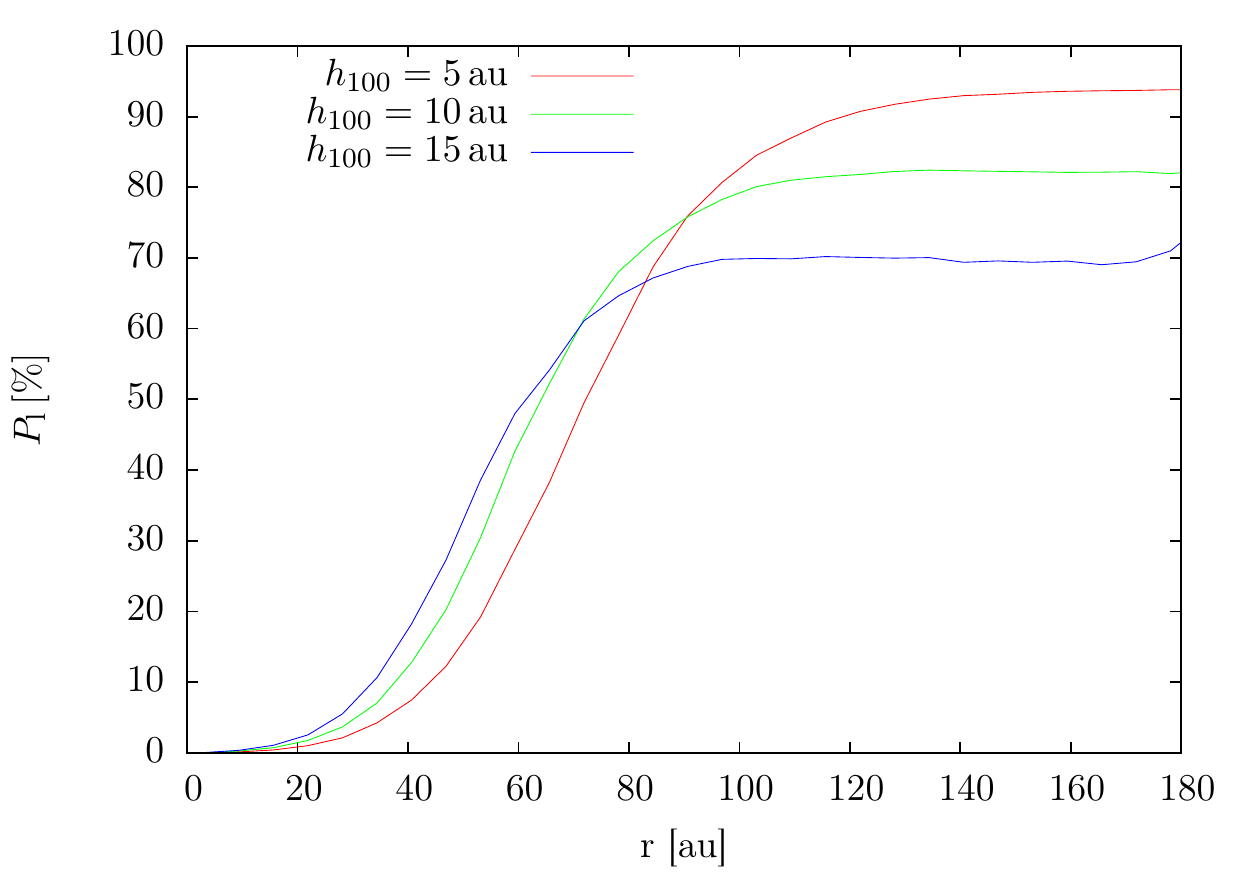}
 \end{Large}
 \caption{Comparison of the radial polarization profiles for disks with various scale heights.
  For details, see Sect. \ref{sec:param}.}
 \label{img:vergleich_h}
\end{figure}

\textbf{Scale height $h_\text{100}$:} The radial dependence of the linear polarization degree for 
various scale heights $h_\text{100}$ is shown in Fig.\,\ref{img:vergleich_h}.
For radial distances below $\approx70\,\mathrm{au}$, the linear polarization degree increases with increasing
scale height. Further outside, i.e., for radial distances above $\approx90\,\mathrm{au}$, the linear polarization degree
decreases with increasing scale height.
First, a smaller scale height corresponds to a more compact disk in the vertical direction, reducing the width of the layer in which
mid-IR radiation is produced. 
Second, the height of the $\tau=1$-surface is decreased as well, decreasing the deviation of the scattering angle from $90^\circ$.
For smaller radial distances, the first effect is more important, while for larger radial
distances, the second dominates.
For grain size distributions with larger maximum grain radii, the first effect only dominates inside
a radial distance of $\sim 20\,\mathrm{au}$ and the differences between the 
linear polarization degrees for various scale heights decrease.

\textbf{Radial density parameter $\alpha$:} A steeper radial dust profile increases the linear polarization degree slightly.
However, a more detailed analysis is omitted in this case because this effect is negligible compared
to a change in disk flaring or scale height.
Grain size distributions with larger maximum grain radii show the same trends.

In summary,
increasing the flaring exponent decreases the linear polarization degree at smaller radial distances,
while an increasing scale height increases the linear polarization degree further inside and decreases it further outside.

 \subsubsection{Inclination dependence of the polarization pattern}
 \label{sec:polmap}
 
\begin{figure}
  \centering
  \includegraphics[width=1.0\columnwidth]{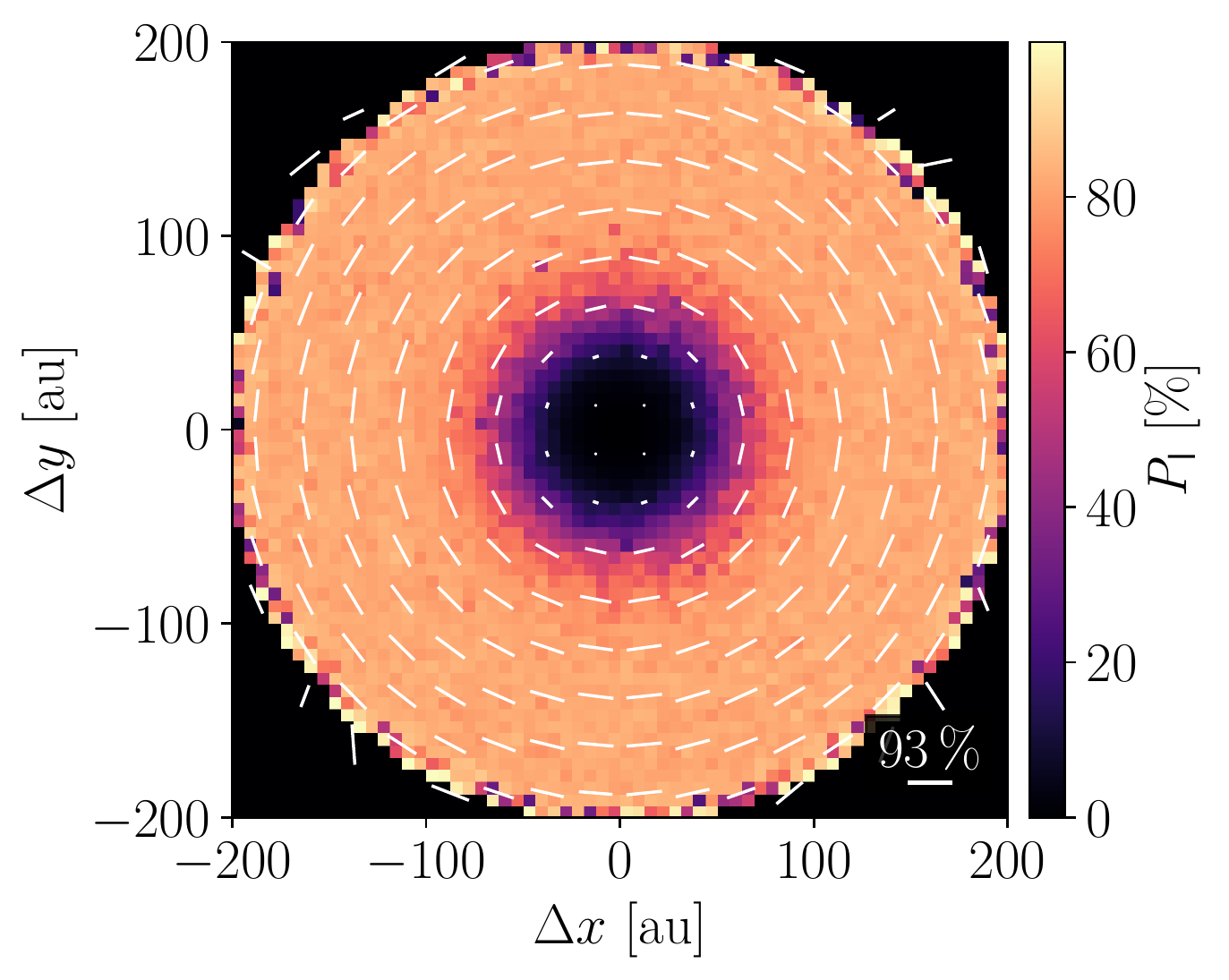}
  \caption{Polarization map for the reference disk.
  For details, see Sect.\,\ref{sec:polmap}.}
  \label{img:polmap}
 \end{figure}

In the preceding chapters, we considered a face-on disk, causing a centro-symmetric
polarization pattern as shown in Fig.\,\ref{img:polmap}.
Before showing how the polarization pattern changes with inclination, we briefly discuss 
 the change of the net linear polarization degree with inclination. 
 This value is zero for a face-on disk owing to the centro-symmetry
 of the polarization pattern. With increasing inclination, the net linear polarization degree 
 increases because the polarization pattern gets asymmetric.
 However, this is difficult to observe because, below an inclination of $\sim 70^\circ$, 
 the net linear polarization degree is very low ($< 1\,\mathrm{\%}$).
 For inclinations above $\sim 70^\circ$, the net linear polarization degree is larger
 but the net mid-infrared flux, resulting from scattering in the upper disk layers only
 becomes very low.
 
 

\begin{figure}
  \centering
   \begin{subfigure}[b]{0.51\columnwidth}
 \includegraphics[width=1.0\textwidth]{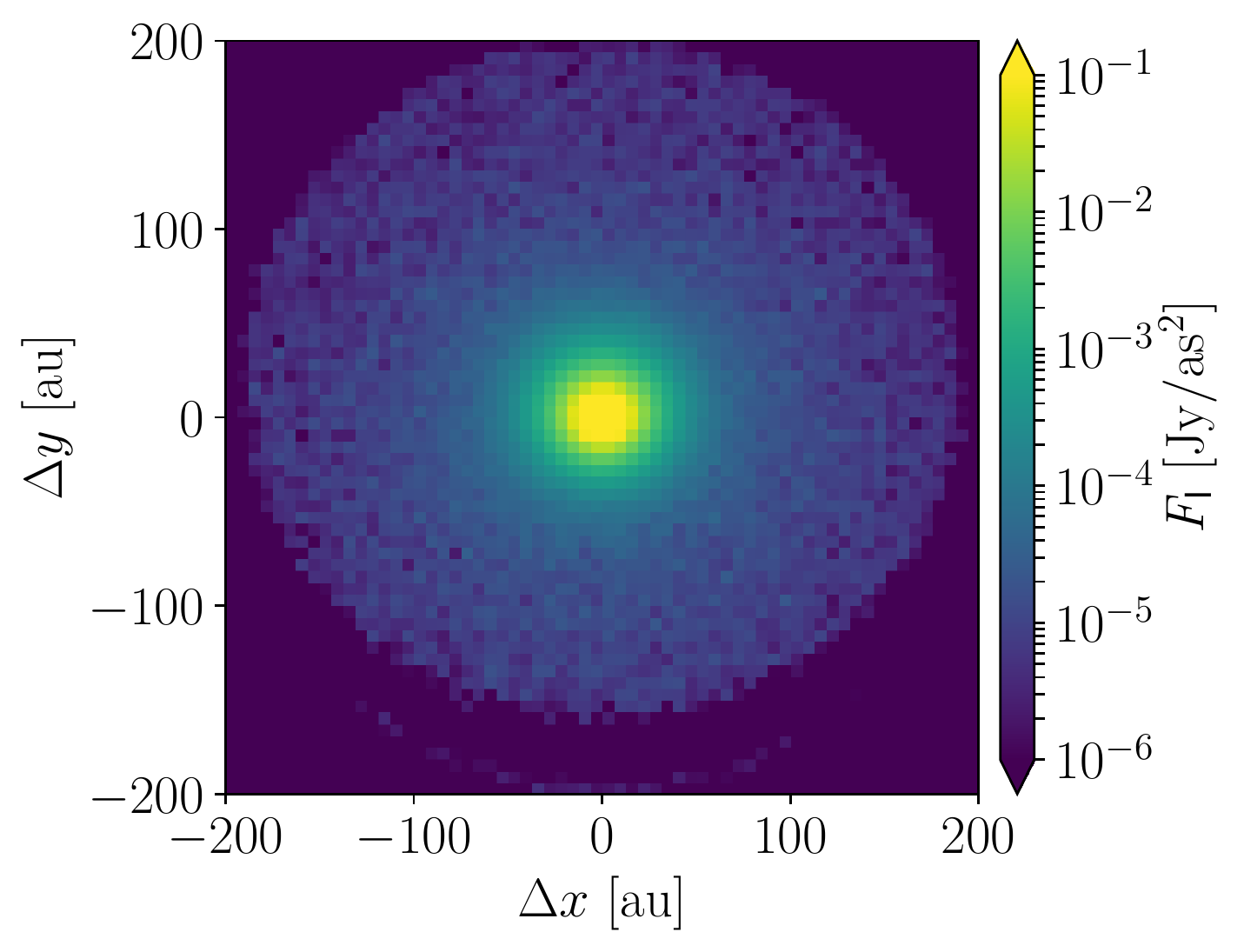}
\end{subfigure}
  \begin{subfigure}[b]{0.48\columnwidth}
 \includegraphics[width=1.0\textwidth]{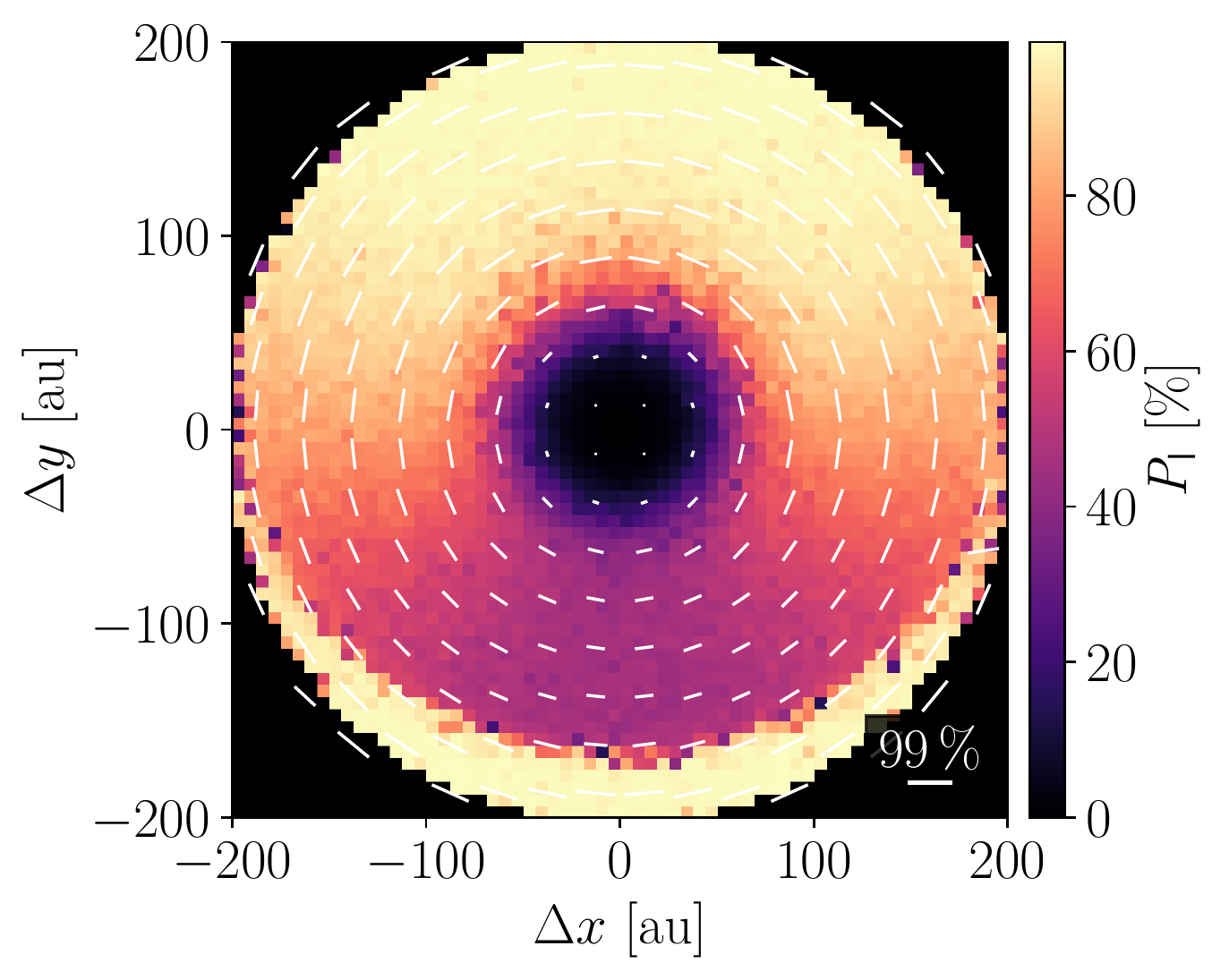}
\end{subfigure}
 \begin{subfigure}[b]{0.51\columnwidth}
 \includegraphics[width=1.0\textwidth]{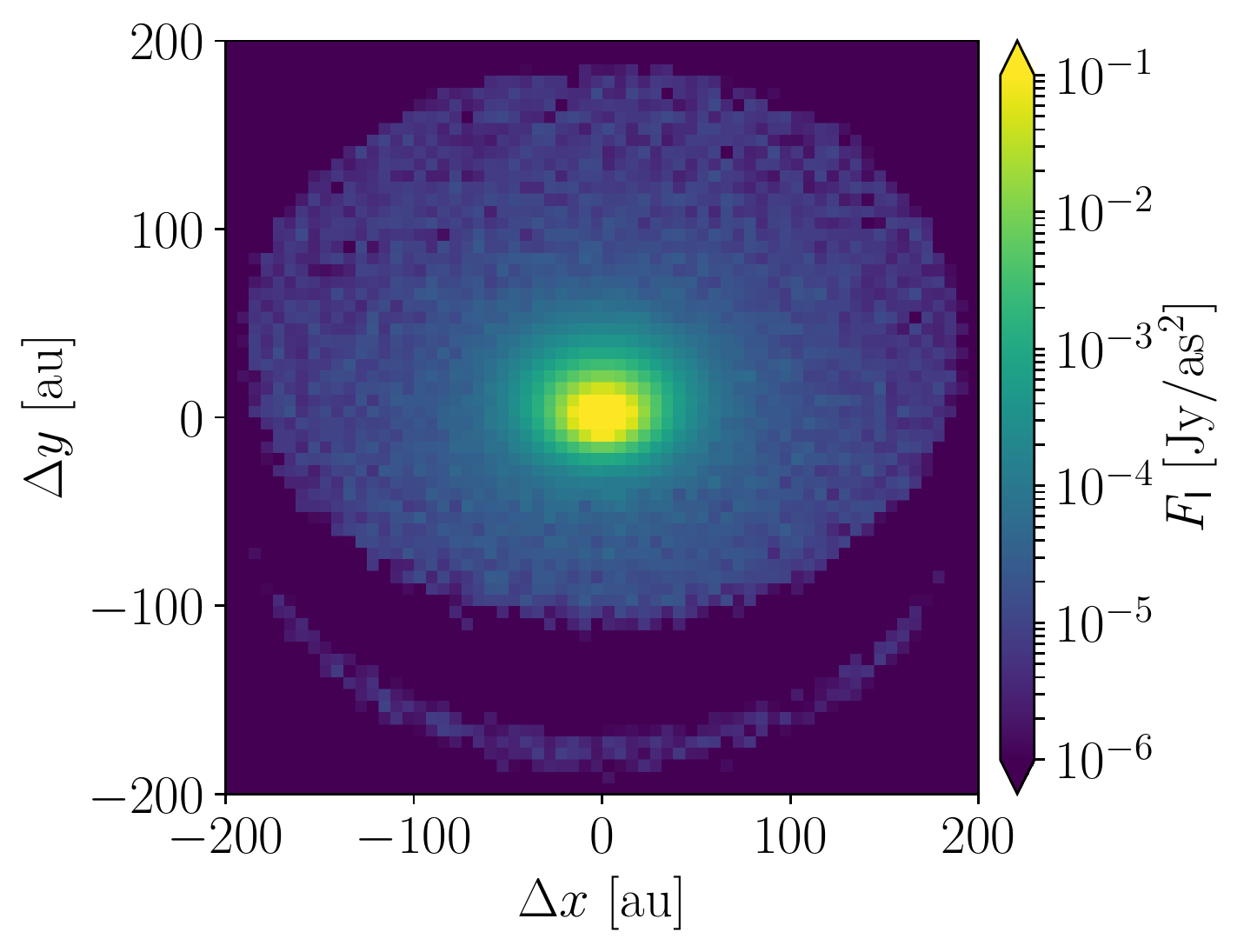}
\end{subfigure}
\begin{subfigure}[b]{0.48\columnwidth}
  \includegraphics[width=1.0\textwidth]{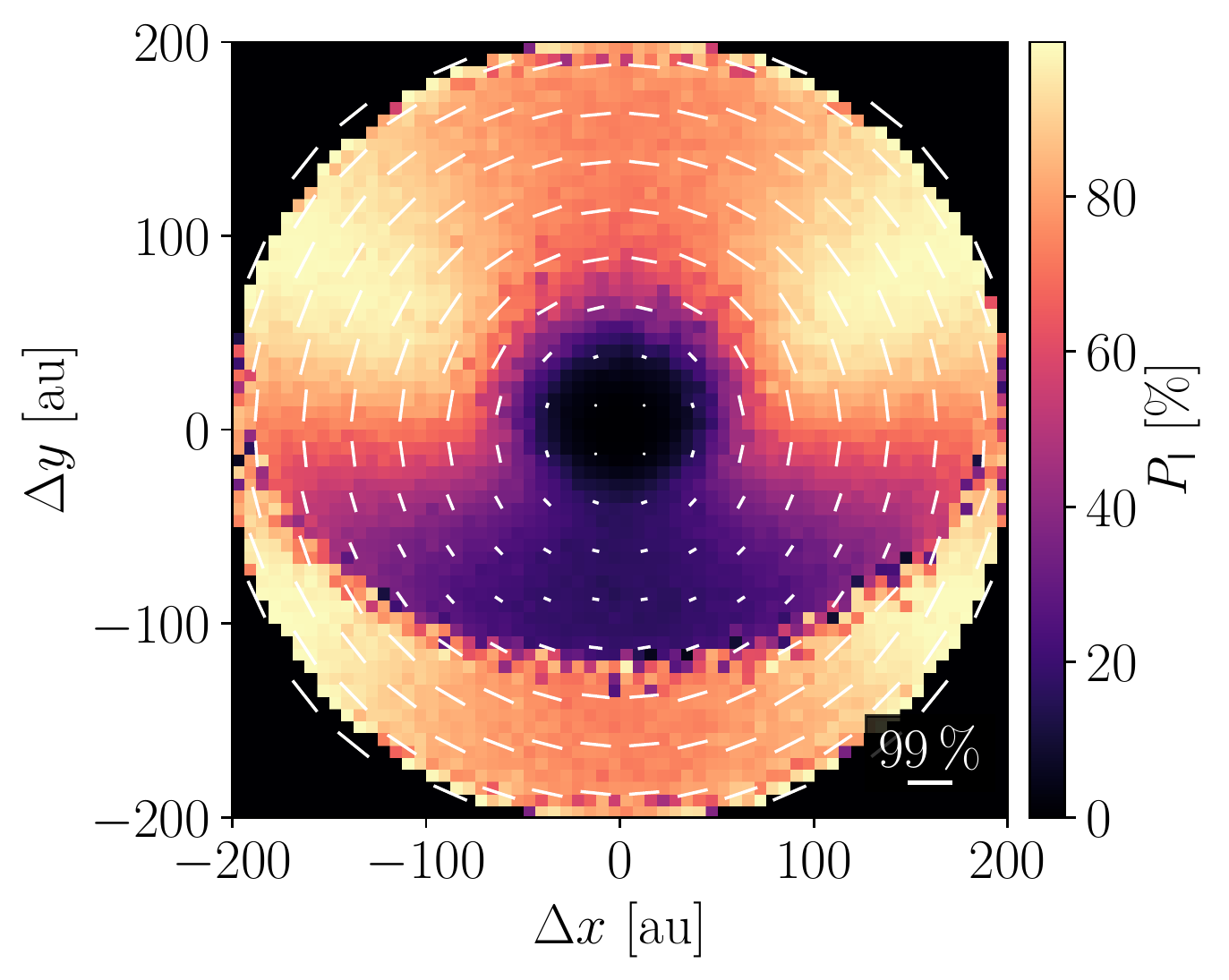}
\end{subfigure}
 \begin{subfigure}[b]{0.51\columnwidth}
 \includegraphics[width=1.0\textwidth]{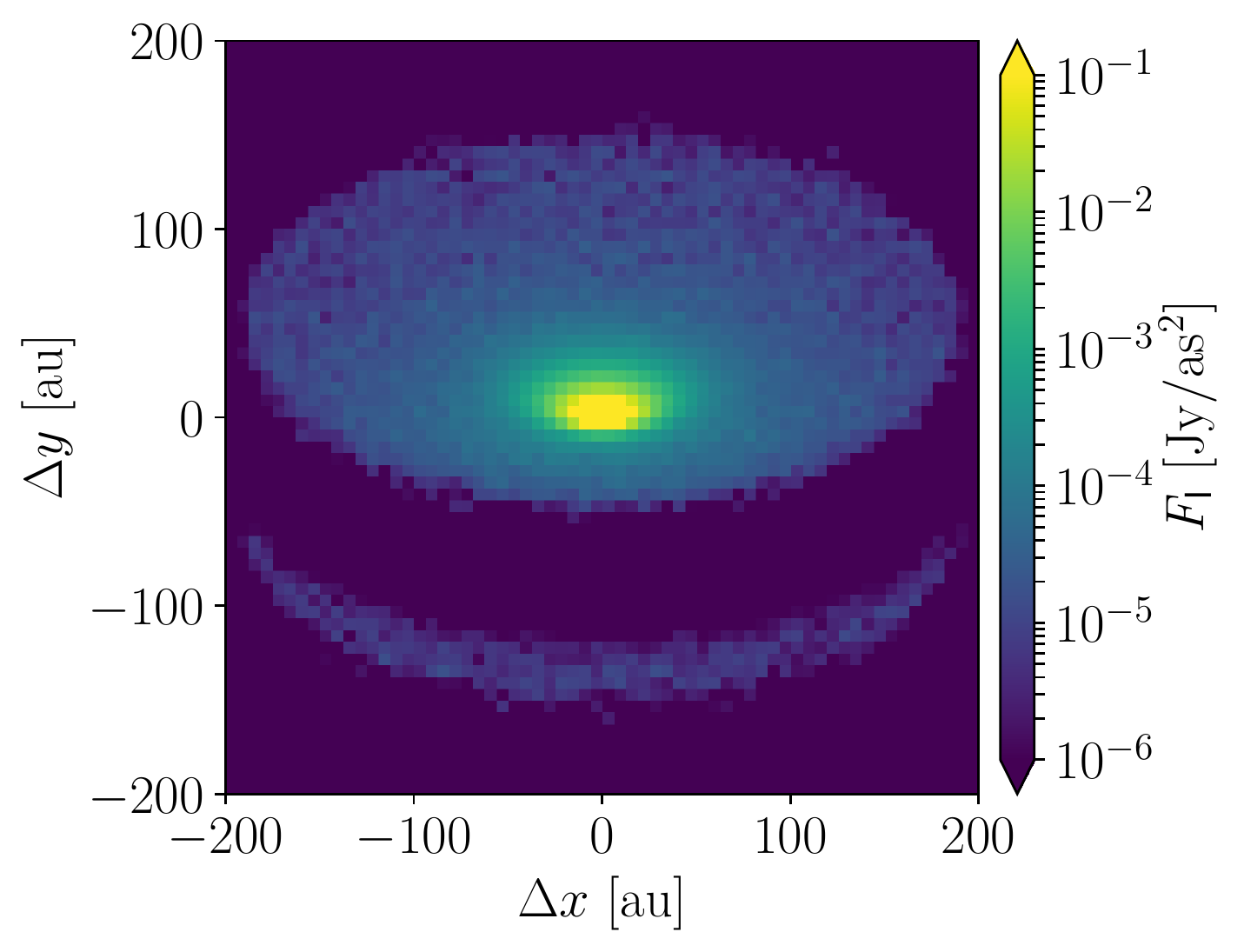}
\end{subfigure}
\begin{subfigure}[b]{0.48\columnwidth}
  \includegraphics[width=1.0\textwidth]{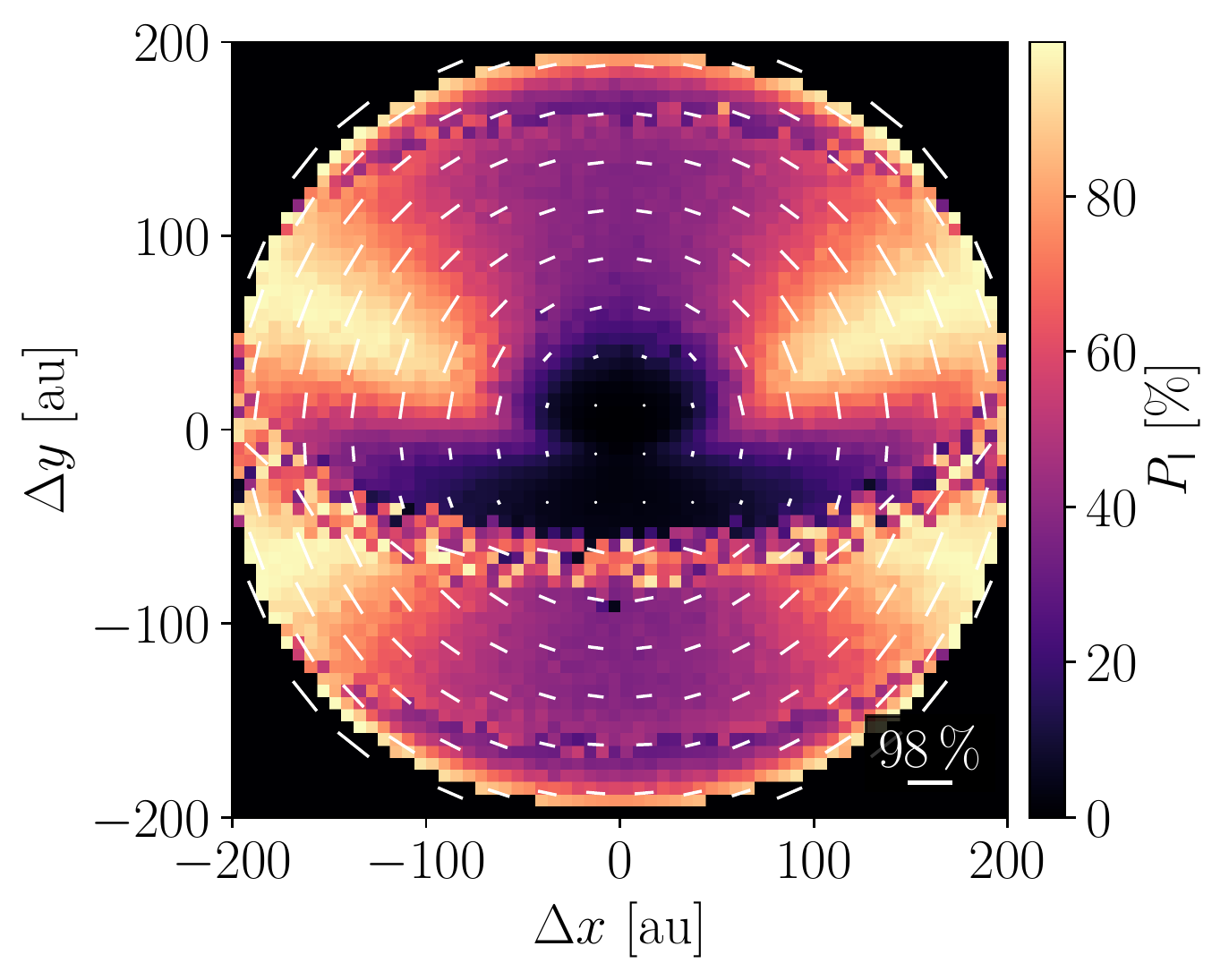}
\end{subfigure}
\caption{Intensity maps (left) and polarization maps (right) for an inclination of $20^\circ$ (top), $40^\circ$ (center) and $60^\circ$ (bottom). 
 For details, see Sect.\,\ref{sec:polmap}.}
\label{img:mc_incl}
 \end{figure}

 We now show the polarization maps for three different inclinations in Fig.\,\ref{img:mc_incl}, right.
The part of the disk that is further away from the observer ("far edge" in the following) is shown
 in the upper half of each polarization map, while the part that is closer to the observer ("close edge") is shown in the lower half.

For an inclination of $20\,\mathrm{^\circ}$, the linear polarization degree for the close edge of the disk is decreased compared to the 
face-on disk because an increasing fraction of the radiation is scattered at angles $\gtrless 90^\circ$.
The linear polarization degree for the far edge is increased because the 
scattering angle is closer to $90^\circ$.
At the same time, the orientation of the polarization vectors remains unchanged.
Increasing the disk inclination, the linear polarization degree for the close edge decreases with the exception of a band in the lower part of the 
map that is produced by the flared outer rim of the disk. This band has a high linear polarization degree, but it is difficult to observe because
of the low flux density (Fig.\,\ref{img:mc_incl}, left).
The linear polarization degree of the far edge decreases as well
because the scattering angle deviates from perpendicular again.
However, the distribution of the directions of the polarization vectors remains qualitatively unchanged.

\begin{figure}
  \centering
  \begin{subfigure}[b]{0.5\columnwidth}
 \includegraphics[width=1.0\textwidth]{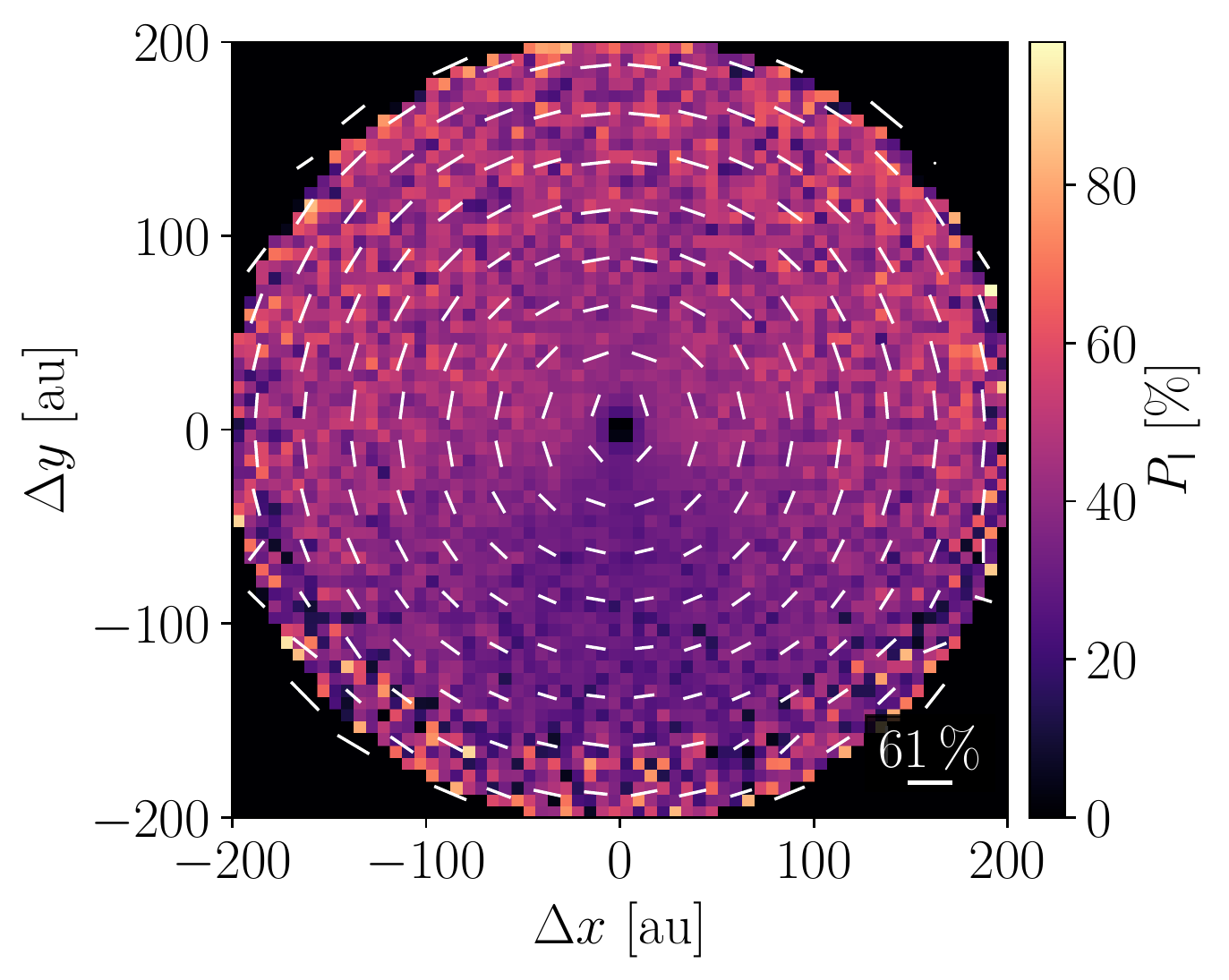}
\end{subfigure}
\begin{subfigure}[b]{0.5\columnwidth}
  \includegraphics[width=1.0\textwidth]{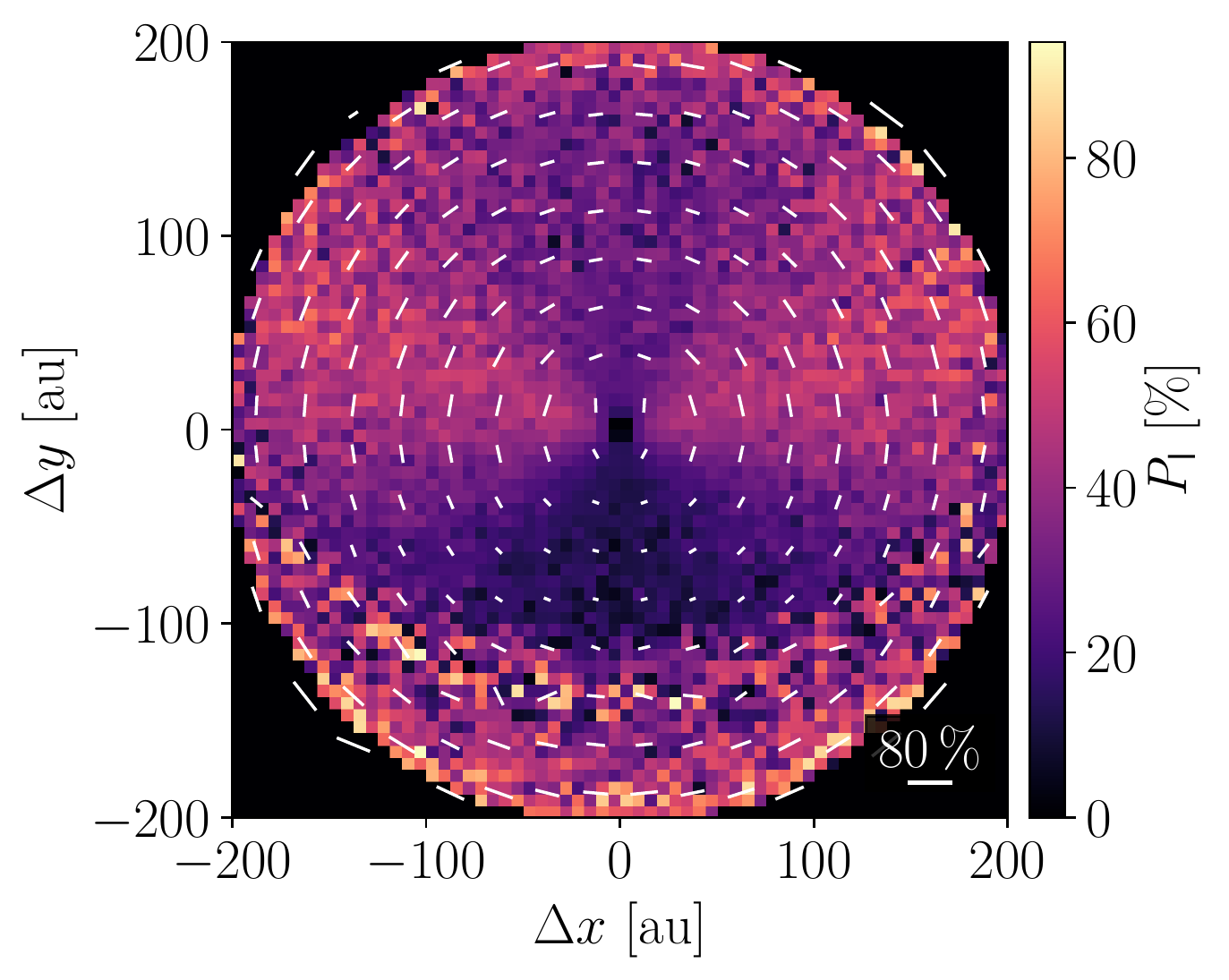}
\end{subfigure}
\begin{subfigure}[b]{0.5\columnwidth}
  \includegraphics[width=1.0\textwidth]{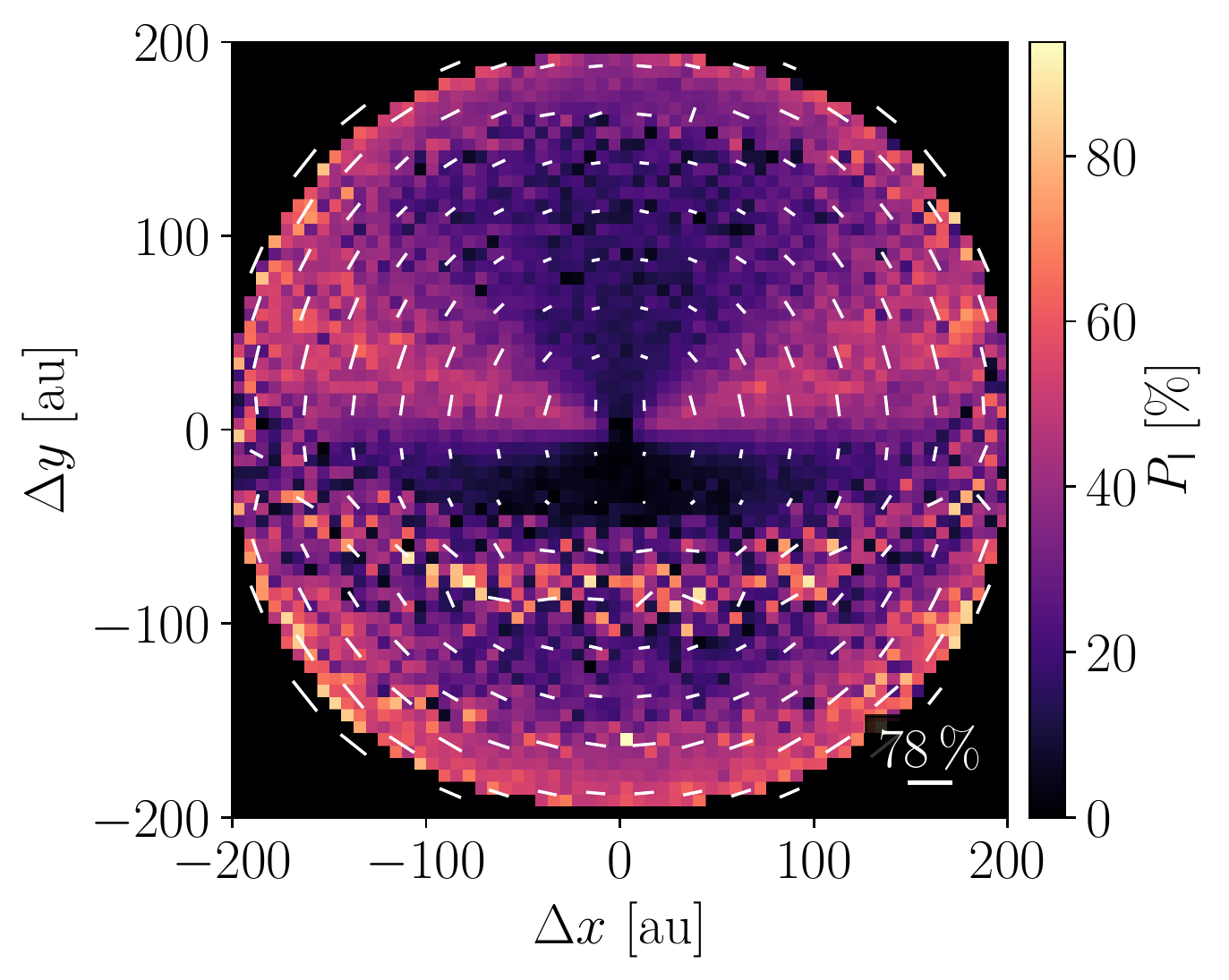}
\end{subfigure}
\caption{Polarization maps for a grain size distribution with a maximum grain radius of $2.5\,\mathrm{\mu\meter}$
and an inclination of $20^\circ$ (left), $40^\circ$ (center) and $60^\circ$ (right). 
 For details, see Sect.\,\ref{sec:polmap}.}
\label{img:mc_incl_mrn10}
 \end{figure}

For grain size distributions with larger maximum grain radii,
the polarization degree is decreased, but the overall polarization pattern is unchanged.
This is illustrated in Fig.\,\ref{img:mc_incl_mrn10} for $a_\text{max}=2.5\,\mathrm{\mu\meter}$.
 
 In summary, increasing the disk inclination increases the net linear polarization degree. 
 The polarization maps become asymmetric with increasing inclination, and the linear polarization degree, especially of the close edge,
 is decreased. However, the distribution of the directions of the polarization vectors remains qualitatively unchanged.
 
 \subsubsection{Herbig Ae star}
 \label{sec:vergleich_herbig}
 
\begin{figure}
  \centering
\begin{Large}
  \includegraphics[width=1.0\columnwidth]{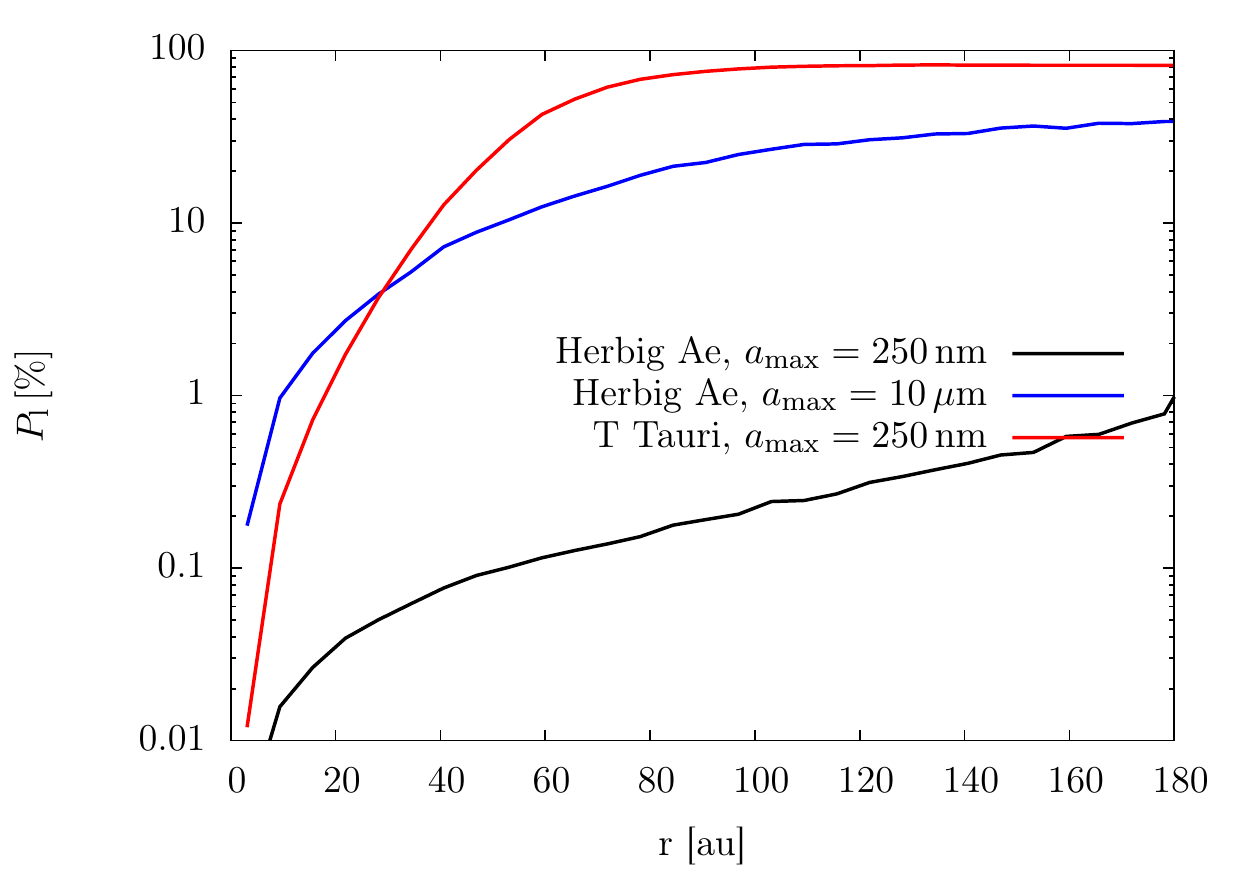}
  \end{Large}
\caption{Comparison of the radial polarization profiles for disks around the considered T Tauri and a Herbig Ae star. 
For details, see Sect.\,\ref{sec:vergleich_herbig}.}
\label{img:vergleich_herbig_wl}
 \end{figure}

 Herbig Ae stars are significantly more luminous than T Tauri stars. Hence, the hotter
 disks are stronger
 and larger mid-IR radiation sources than their T Tauri counterparts. 
 At the same time, unscattered thermal re-emission radiation is
 expected to dominate over the scattered radiation out to larger radial distances. This lowers the expected linear polarization degree
 at a given radial distance from the central star.
 To verify this hypothesis, we computed the radial polarization profiles for a disk heated by a Herbig Ae star and grain species
 with multiple maximum grain radii and show these in Fig.\,\ref{img:vergleich_herbig_wl}. The polarization profile for $a_\text{max}=2.5\,\mathrm{\mu\meter}$
 is omitted because the results are similar to the case of $a_\text{max}=10\,\mathrm{\mu\meter}$.
 
\begin{figure}
  \centering
  \begin{Large}
  \includegraphics[width=1.0\columnwidth]{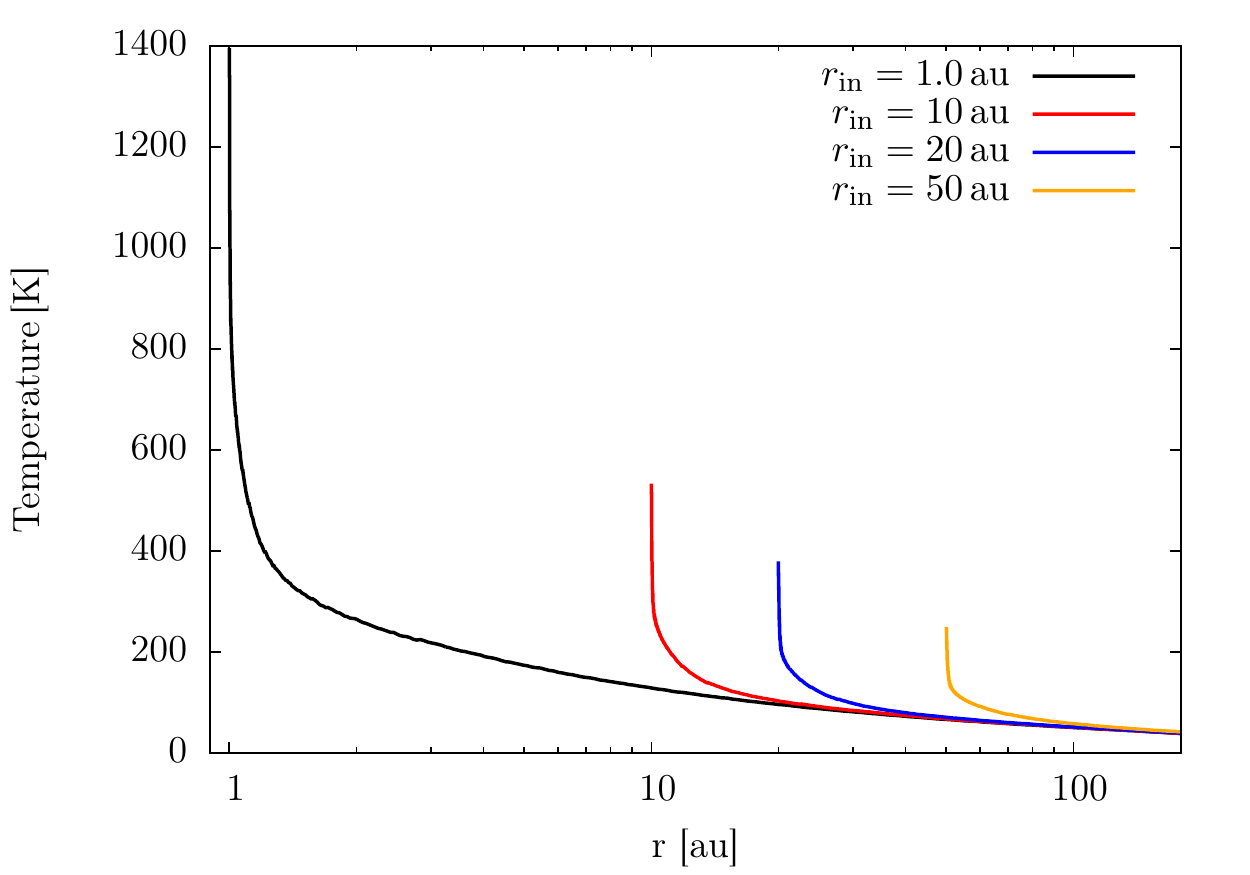}
  \end{Large}
  \caption{Comparison of the temperatures at the $\tau(9.5\,\mathrm{\mu\meter})=1$ surface as seen from above the disk, perpendicular to the disk midplane,
  vs. radial distance for multiple inner radii and a Herbig Ae star
  as central star.
  For details, see Sect.\,\ref{sec:vergleich_herbig}.}
  \label{img:Innenradius_herbig_temp}
 \end{figure}

 As shown in Fig. 23,
 the mid-IR radiation from disks heated by a Herbig Ae star is almost unpolarized for grain size distributions with small maximum grain radii.
 For grain size distributions with larger maximum grain radii, the scattered radiation is much stronger. Therefore, the radiation of the disk is
 significantly polarized even if the central star is a Herbig Ae star.
 The other trends are similar to those for the T Tauri star. Only the increase of the linear polarization degree happens at the smallest radial
 distance for an inner radius of $\sim20\,\mathrm{au}$ because
 the size of the effective mid-IR emitting region is maximized for this inner radius
 (see Fig.\,\ref{img:Innenradius_herbig_temp}). For grains with larger maximum grain radii ($2.5\,\mathrm{\mu\meter}$ and $10\,\mathrm{\mu\meter}$), the linear polarization degree increases 
 at the smallest radial distance for an inner radius of
 $\sim10\,\mathrm{au}$.
 
 In conclusion, replacing the T Tauri star by a Herbig Ae star substantially increases the flux density but decreases the linear polarization degree
 for grain size distributions with small maximum grain radii.
 The inner radius at which the increase of the linear polarization degree happens furthest inside is $\sim20\,\mathrm{au}$ for the stellar parameters provided in this work
 and $\sim10\,\mathrm{au}$ for grain size distributions with a larger maximum grain radius.
 The other results are similar to those for disks heated by a T Tauri star.
 
 \section{Discussion}
 
 \label{sec:comparison}
 
 
This study was motivated by the fact that the analysis of the linear continuum
polarization potentially allows us to assess the magnetic field  of a protoplanetary
disk. While this assumption is based on the impact of magnetically aligned grains on the 
polarization state of the observable radiation, the role of scattering of radiation
(stellar, thermal (dust) re-emission) should be known.
 In the following, we discuss our results for scattered light polarization in view of their
 potential to distinguish between both polarization mechanisms:
 \begin{enumerate}
 \item For scattering, the linear polarization degree increases with radial distance. 
 In contrast, the linear polarization degree in the case of dichroic emission and absorption shows a weaker and less distinct dependence on
 radial distance (\citealt{2016ApJ...832...18L}).
 \item The linear polarization degree decreases with increasing wavelength for grains with small maximum grain radii ($250\,\mathrm{\nano\meter}$). 
 For larger maximum grain radii ($2.5\,\mathrm{\mu\meter}$, $10\,\mathrm{\mu\meter}$), it converges to the final value more slowly,
 but this final value depends on both the wavelength and maximum dust grain radius.
 In the context of the disk and dust model investigated by \cite{2004MNRAS.348..279A}, the authors found that the linear
 polarization degree has a maximum at a wavelength between $10\,\mathrm{\mu\meter}$ and $11\,\mathrm{\mu\meter}$
 and decreases for longer wavelengths.
 However, at wavelengths above the \textit{N} band, it is possible that the linear polarization degree increases
 with increasing wavelength (\citealt{2007ApJ...669.1085C}). Using this trend to distinguish scattering polarization from polarization due 
 to dichroic emission and absorption requires independent measurements of the dust grain properties (e.g., composition).
 \item The polarization vectors for polarization by scattering show a tangential pattern, while the polarization 
 vectors due to dichroic emission and absorption depend on the alignment of the dust grains (\citealt{0004-637X-839-1-56}).
 This alignment depends on the magnetic field. Therefore, the polarization vectors can also be used to obtain information
 about the magnetic field structure. However, if the dust grains are magnetically poor and align only with the radiation direction, 
 the polarization vectors can be tangential even if the polarization is caused by dichroic emission and absorption. Therefore,
 an analysis of the polarization vectors alone does not always yield distinct results.
 \end{enumerate}
 
 In summary, polarization due to dichroic emission and absorption and scattering shows different trends, potentially allowing us to
 distinguish both
 polarization mechanisms. However,  we recommend using multiple trends to make distinguishing
 the polarization mechanisms more clear because of the potential for ambiguity.
 
 \section{Conclusions}
 \label{sec:conclusion}
 
 In the first part of our mid-IR study (Sect.\,\ref{sec:ratio}), we computed the ratio of thermal re-emission radiation
 to scattered stellar radiation
  to determine the dominating radiation source. 
 We find that thermal re-emission radiation is stronger than scattered stellar radiation
 for disks with an inner radius below $\sim10\,\mathrm{au}$, a T Tauri star as the central star
 and a wavelength of $9.5\,\mathrm{\mu\meter}$.
 Therefore, self-scattering potentially has a greater influence on the resulting linear polarization degree
 than scattered stellar radiation. However, the scattered stellar radiation may become significant
 in the case of transition disks with sufficiently large inner radii.
 Using a Herbig Ae star as central star significantly increases the thermal re-emission radiation and therefore also the flux ratio. 
 
 In the second part, we performed studies of the polarization due to scattering in the mid-IR. The following trends were identified based on these studies:
 \begin{enumerate}
 \item For scattered stellar radiation only, the linear polarization degree slowly decreases with radial distance (Sect.\,\ref{sec:quelle}).
  \item For thermal re-emission radiation and for combined thermal re-emission radiation and scattered stellar radiation,
  the linear polarization degree increases with radial distance (Sect.\,\ref{sec:quelle} and \ref{sec:vergleich_wl}). In contrast,
  the linear polarization degree due to dichroic emission and absorption shows less distinct and less strong dependence on
  radial distance (\citealt{2016ApJ...832...18L}).
  \item The linear polarization degree decreases for longer wavelengths and grain size distributions with small maximum grain radii ($250\,\mathrm{\nano\meter}$).
  For grain size distributions with larger maximum grain radii ($2.5\,\mathrm{\mu\meter}$, $10\,\mathrm{\mu\meter}$), this degree converges slower.
  The polarization degree at large radial distances from the star depends on both the wavelength and the maximum dust grain radius (Sect.\,\ref{sec:vergleich_wl}).
  Using this trend to distinguish scattering polarization from polarization due to dichroic emission and absorption
  requires using an independent method to assess the dust grain properties (e.g., composition).
  \item The linear polarization degree increases further inside for larger dust masses, while
  the maximum linear polarization degree decreases (Sect.\,\ref{sec:vergleich_mass}).
  For grain size distributions with the largest maximum dust grain radius, the linear polarization
  degree increases with dust mass for all radial distances.
  \item The linear polarization degree increases at the smallest radial distance
  if the size of the effective radiation source is maximized. For smaller and larger corresponding
  inner radii, the linear polarization degree increases further outside (Sect.\,\ref{sec:vergleich_ri}).
  For grain size distributions with larger maximum dust grain radii, the maximum polarization degree
  depends on the inner disk radius.
  \item An increase of the flaring exponent $\beta$ decreases the linear polarization degree inside a radial distance of
  $\approx100\,\mathrm{au}$ (grain size distributions with small maximum grain radii) and $\approx 20\,\mathrm{au}$ (grain size distributions with
  larger maximum grain radii; Sect.\,\ref{sec:param}).
  \item An increase of the scale height $h_\text{100}$ increases the linear polarization degree inside a radial
  distance of $\approx70\,\mathrm{au}$ for grain size distributions with small maximum grain radii
  and $\approx 20\,\mathrm{au}$ for grain size distributions with
  larger maximum grain radii. 
  Further outside, at radial distances above $\approx90\,\mathrm{au}$ 
 for grain size distributions with small maximum grain radii and $\approx30\,\mathrm{au}$
  for grain size distributions with
  larger maximum grain radii, an increasing scale height 
  decreases the linear polarization degree (Sect.\,\ref{sec:param}).
  \item The net linear polarization degree increases with disk inclination (Sect.\,\ref{sec:polmap}).
  \item The polarization vectors show a tangential pattern (Sect.\,\ref{sec:polmap}).
  This distribution of directions remains qualitatively unchanged for inclined disks.
  For polarization due to dichroic emission and absorption, the polarization vectors depend on the alignment
  of the dust grains (\citealt{0004-637X-839-1-56}). Hence, if the magnetic field is strong enough,
  the polarization vectors significantly differ from a tangential pattern.
  \item With increasing inclination, the linear polarization degree at the close edge of the polarization map 
  is decreased, while that at the far edge is first increased and then decreased (Sect.\,\ref{sec:polmap}).
  \item Using a Herbig Ae star instead of a T Tauri star as central star significantly decreases the polarization
  degree for grain size distributions with small maximum grain radii. For larger maximum grain radii, the effect is less 
  great (Sect.\,\ref{sec:vergleich_herbig}).
 \end{enumerate}

 
 Three of these trends (items 2, 3, and 9) potentially enable us to distinguish polarization due to scattering from polarization due to 
  dichroic emission and absorption. However, we recommend using multiple trends to distinguish the polarization
  mechanisms more clearly.
 Using these trends requires a sufficient spatial resolution to map structures on sub-disk scales ($\approx 50\,\mathrm{au}$).
 Additionally, using item 3 requires an independent method to constrain the dust grain properties.
 
 
 \begin{acknowledgements}
 The authors acknowledge support by the DFG program WO857/18-1.
\end{acknowledgements}

%
  \bibliographystyle{aa} 
  \bibliography{paper} 
%

\end{document}